\documentclass{article}

\usepackage{amsmath,amsfonts,bm}

\def\eqref#1{equation~\ref{#1}}

\def\1{\bm{1}}

\DeclareMathAlphabet{\mathsfit}{\encodingdefault}{\sfdefault}{m}{sl}
\SetMathAlphabet{\mathsfit}{bold}{\encodingdefault}{\sfdefault}{bx}{n}

\usepackage{xcolor}
\PassOptionsToPackage{hyphens}{url}
\usepackage[hyperfootnotes=false]{hyperref}
\usepackage[accepted]{mlsys2025} %
\makeatletter\renewcommand{\mlsys@appearing}{Preprint.}\makeatother
\usepackage{url}
\usepackage{graphicx}
\usepackage{array}
\usepackage{booktabs} %
\usepackage{subcaption}
\usepackage{xspace}

\newcommand{\hecbench}{\texttt{HeCBench}\xspace}
\newcommand{\rai}{RAI\xspace}
\newcommand{\bbound}{BB\xspace}
\newcommand{\cbound}{CB\xspace}

\newcommand{\Ovariant}{\texttt{\textbf{O}}\xspace}
\newcommand{\CIvariant}{\texttt{\textbf{CI}}\xspace}
\newcommand{\Rvariant}{\texttt{\textbf{R}}\xspace}
\newcommand{\RDvariant}{\texttt{\textbf{RD}}\xspace}
\newcommand{\zeroshot}{\texttt{ZS}\xspace}
\newcommand{\zssc}{\texttt{ZS-Code}\xspace}
\newcommand{\zsisa}{\texttt{ZS-ISA}\xspace}
\newcommand{\agentic}{\texttt{AG}\xspace}
\newcommand{\toolcall}{\texttt{rai\_prediction\_result}\xspace}
\definecolor{rooflineCB}{HTML}{0072B2}
\definecolor{rooflineBB}{HTML}{D55E00}
\definecolor{rooflineNC}{HTML}{6C6C6C}
\newcommand{\CBmark}{\textcolor{rooflineCB}{\raisebox{0.15ex}{\rule{0.72ex}{0.72ex}}}}
\newcommand{\BBmark}{\textcolor{rooflineBB}{\ensuremath{\circ}}}
\newcommand{\NCmark}{\textcolor{rooflineNC}{--}}
\newcommand{\regimesig}[4]{\mbox{#1\hspace{0.08em}#2\hspace{0.08em}#3\hspace{0.08em}#4}}

\usepackage[most]{tcolorbox}
\newtcolorbox{rqsummary}[1]{enhanced, sharp corners, boxrule=0pt, leftrule=2pt,
  colback=green!8, colframe=green!45!black, left=4pt, right=4pt, top=2pt, bottom=2pt,
  fontupper=\footnotesize, before upper={\textbf{#1 Summary:}\ }}

\newcommand{\deepseekflash}{\texttt{deepseek-v4-flash}\xspace}
\newcommand{\geminiflash}{\texttt{gemini-3.7-flash}\xspace}
\newcommand{\gptsol}{\texttt{gpt-5.6-sol}\xspace}
\newcommand{\deepseekshort}{\texttt{deepseek}\xspace}
\newcommand{\geminishort}{\texttt{gemini}\xspace}
\newcommand{\gptshort}{\texttt{gpt-5.6}\xspace}

\newcommand{\papertitle}{Estimating GPU Roofline Arithmetic Intensity with Off-the-Shelf LLMs and Compiler Artifacts}
\mlsystitlerunning{\papertitle}

\begin{document}
\twocolumn[
\mlsystitle{\papertitle}
\begin{mlsysauthorlist}
\mlsysauthor{Gregory Bolet}{vt}
\mlsysauthor{Giorgis Georgakoudis}{intel}
\mlsysauthor{Konstantinos Parasyris}{intel}
\mlsysauthor{Harshitha Menon}{llnl}
\mlsysauthor{Niranjan Hasabnis}{codemetal}
\mlsysauthor{Kirk W. Cameron}{vt}
\mlsysauthor{Gal Oren}{stanford,technion}
\end{mlsysauthorlist}
\mlsysaffiliation{vt}{Virginia Tech, Blacksburg, Virginia, USA}
\mlsysaffiliation{intel}{Intel Corporation, USA}
\mlsysaffiliation{llnl}{Lawrence Livermore National Laboratory, Livermore, California, USA}
\mlsysaffiliation{codemetal}{Codemetal, Boston, Massachusetts, USA}
\mlsysaffiliation{stanford}{Stanford University, Stanford, California, USA}
\mlsysaffiliation{technion}{Technion, Haifa, Israel}
\mlsyscorrespondingauthor{Gregory Bolet}{gbolet@vt.edu}
\mlsyscorrespondingauthor{Giorgis Georgakoudis}{giorgis.georgakoudis@intel.com}
\mlsyscorrespondingauthor{Konstantinos Parasyris}{konstantinos.parasyris@intel.com}
\mlsyscorrespondingauthor{Harshitha Menon}{gopalakrishn1@llnl.gov}
\mlsyscorrespondingauthor{Niranjan Hasabnis}{niranjan@codemetal.ai}
\mlsyscorrespondingauthor{Kirk W. Cameron}{cameron@vt.edu}
\mlsyscorrespondingauthor{Gal Oren}{galoren@stanford.edu}
\vskip 0.3in
\begin{abstract}
GPU developers use profiling to understand computation and data movement, but access to target hardware can delay optimization.
We investigate whether off-the-shelf large language models (LLMs) can estimate Roofline arithmetic intensity (the ratio of floating-point operations (FLOPs) to device-memory bytes) without executing the target GPU workload.
We evaluate three models (\deepseekflash, \geminiflash, and \gptsol) across 179 combinations of five kernel families, four NVIDIA and AMD GPUs, four parallel runtimes (CUDA, HIP, OpenMP, and OpenCL), and modified code variants, comparing source-only queries, queries with compiler artifacts, and sandboxed agentic analysis.
Compiler artifacts raise the share of arithmetic-intensity estimates within a factor of 1.20 of profiled values from 8--14\% to 26--32\%, while the resulting bandwidth-bound or compute-bound classifications are correct in 78--92\% of runs.
With compiler artifacts, FLOP estimates meet the $1.20\times$ tolerance in 59--88\% of runs, compared with 36--40\% for device-memory traffic.
Agentic analysis increases the share of arithmetic-intensity estimates within the $1.20\times$ tolerance by 6.5 percentage points for \gptsol, at $3.6\times$ the median cost and $2.5\times$ the median latency of compiler-assisted queries; the other models incur higher costs without a statistically clear improvement.
These findings suggest LLM estimates could help prioritize optimization when GPU access is constrained, while identifying cache-dependent device-memory traffic as the main obstacle to precise arithmetic-intensity prediction.
\end{abstract}
]
\printAffiliationsAndNotice{}
\section{Introduction}
\label{sec:introduction}

\suppressfloats[t]

\begin{figure*}[t]
    \centering
    \includegraphics[width=0.75\linewidth]{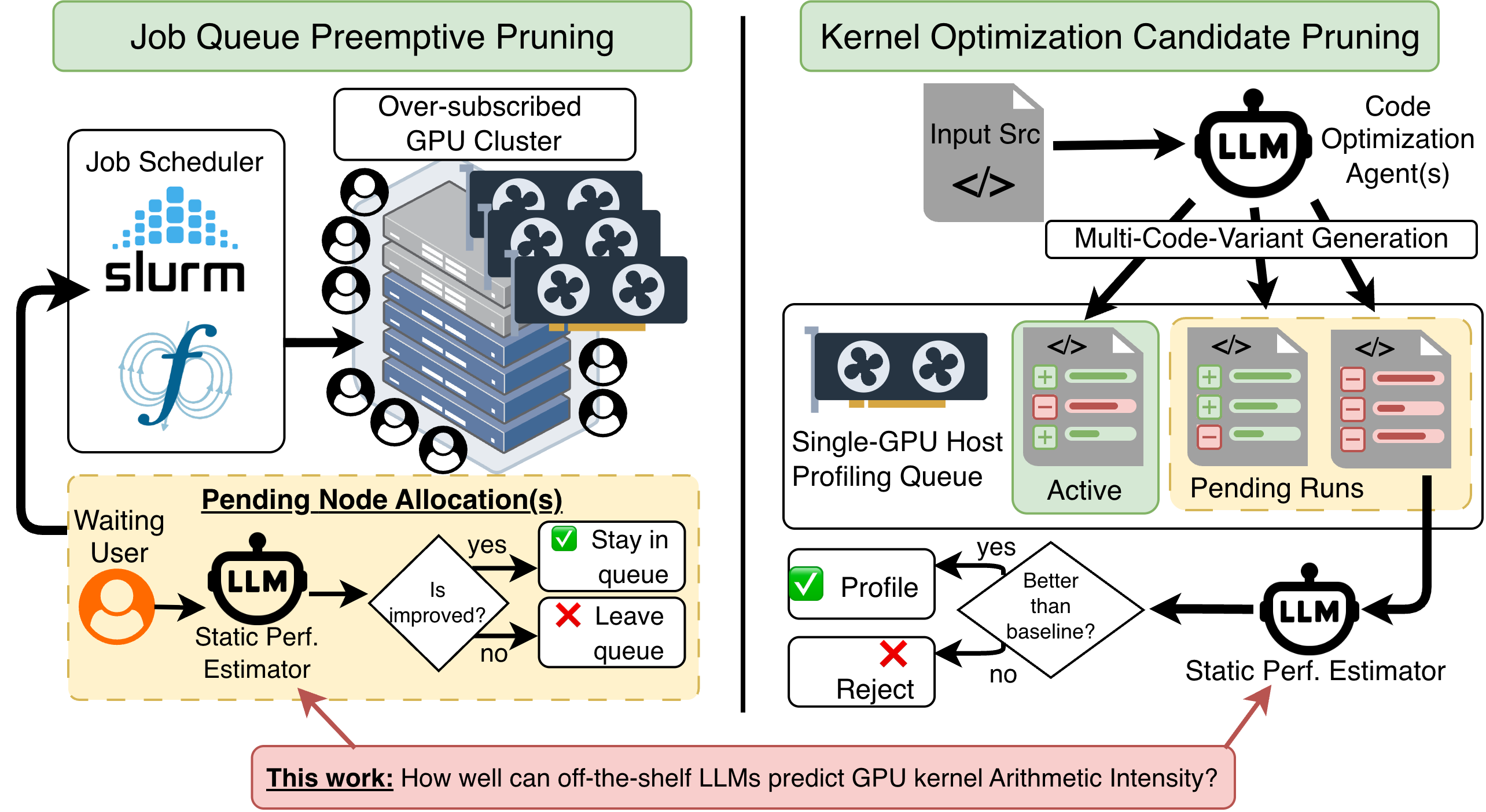}
    \caption{Potential uses of static estimates when GPU access is constrained: screening jobs awaiting cluster allocation (left) and prioritizing generated optimization candidates for profiling (right).
    We evaluate the underlying arithmetic-intensity predictions.}
    \label{fig:intro-motivation}
\end{figure*}

Which should a developer optimize first: a GPU kernel's computation or its data movement?
Profiling the kernel on its target graphics processing unit (GPU) informs this decision, but access may be delayed by shared-cluster allocations or queues of optimization candidates.
Large language model (LLM) optimization systems also generate candidates that require hardware evaluation \citep{zhang2025cudaforge,nichols2025integrating}.
When candidates are profiled in isolation to avoid interference in device-memory traffic measurements, they must take turns on the target GPU \citep{nvidiaNsightComputeKnownIssues2026}.
LLM assistants are increasingly part of software-development workflows \citep{nichols2024pareval}, making static estimation a potential additional use for tools developers already query.
Figure~\ref{fig:intro-motivation} illustrates how estimates could help prioritize optimization candidates while GPU access is constrained.

Roofline arithmetic intensity (RAI) is floating-point operations (FLOPs) divided by device-memory bytes transferred \citep{williams2009roofline}.
Comparison with the target GPU's balance point identifies the Roofline model's bandwidth-bound or compute-bound regime.
The numerical value also helps bound attainable throughput and shows proximity to the balance point, information a regime label alone cannot provide.
Predicting RAI requires accounting for compiler transformations in executed arithmetic and for access patterns, cache reuse, and the boundary across which memory traffic is measured.
These estimates can guide optimization priorities; execution measurements validate performance improvements.

Analytical and static GPU methods estimate performance from program and architectural information \citep{hong2009analytical,emonds2023cudasap}.
Prior work also explores LLM-based Roofline classification \citep{bolet2025can}, FLOP-count prediction \citep{bolet2025countingrunningevaluatingllms}, and trained numerical GPU performance prediction \citep{nguyennhat2024llmperf,wang2025omniwise}.
Beyond regime classification, we assess how accurately off-the-shelf LLMs recover numerical RAI and its FLOP and device-memory traffic components across targets, and what compiler evidence and sandboxed analysis add without task-specific training.

We compare three strategies without executing the target GPU workload: zero-shot queries using source alone, zero-shot queries supplemented with post-compilation instruction set architecture (ISA) artifacts, and sandboxed agentic analysis.
The study evaluates three models (\deepseekflash, \geminiflash, and \gptsol) across 179 combinations of kernel, variant, GPU, and runtime: five \hecbench kernel families \citep{jin2023hecbench}, four NVIDIA/AMD GPUs, and CUDA, HIP, OpenMP, and OpenCL where supported.
Four variants probe sensitivity to names, compiler settings, inputs, and datatypes as a best-effort check of reliance on recognizable code or recalled measurements from the original source.
Models receive execution inputs and target specifications; profiled counters are withheld for scoring, and failed or incomplete predictions count as misses.
The study distinguishes numerical accuracy from regime accuracy, diagnoses errors in FLOP and device-memory traffic estimates, and measures the cost and latency of additional analysis.
We organize the findings around four research questions:

\begin{enumerate}
    \item[\textbf{RQ1.}] \textbf{Can LLMs predict the \rai of GPU kernels from static evidence?}
    LLMs estimate RAI within $1.20\times$ of the reference in 26--32\% of runs with ISA, up from 8--14\% with source alone; at $2\times$, ISA accuracy reaches 48--60\%.
    With ISA, FLOP accuracy within $1.20\times$ is 59--88\%, versus 36--40\% for device-memory traffic, the recurring numerical weak point.

    \item[\textbf{RQ2.}] \textbf{Can predicted \rai correctly identify a GPU kernel's Roofline regime?}
    Although numerical estimates often miss the $1.20\times$ target, source plus ISA classifies 78--92\% of runs correctly, exceeding a predictor that chooses the more common reference class for each GPU--precision pair (73.2\%, fitted and scored on the same cases).
    For the 40 cases within $2\times$ of the balance point by profiled RAI, ISA raises \gptsol's classification accuracy from 42.5\% to 79.5\%.

    \item[\textbf{RQ3.}] \textbf{Does numerical \rai accuracy vary across workloads, target GPUs, and code variants?}
    Accuracy varies across workload--GPU combinations, consistent with differences in access patterns and effective reuse; ISA gains extend across all four variants.
    For example, \gptsol with ISA reaches 90--100\% accuracy within $1.20\times$ for \texttt{lavaMD} on A100/H100, but 0\% for \texttt{particlefilter} on every GPU.

    \item[\textbf{RQ4.}] \textbf{Can sandboxed analysis improve numerical \rai accuracy beyond zero-shot ISA prompting?}
    Sandboxed analysis improves \gptsol's accuracy within $1.20\times$ by 6.5 percentage points at $3.6\times$ the median cost of source-plus-ISA queries; for the other models, gains remain uncertain, median costs are $6.7$--$17.7\times$ as high, and completion falls as low as 64\%.
\end{enumerate}

\section{Background and Related Work}
\label{sec:background}

\paragraph{Roofline background.}
The Roofline model bounds attainable throughput by the minimum of peak compute throughput and peak memory bandwidth multiplied by arithmetic intensity \citep{williams2009roofline}.
Comparing intensity with the balance point identifies the model's bandwidth-bound or compute-bound regime and guides optimization priorities.
Cache-aware Roofline modeling extends this analysis across the GPU memory hierarchy \citep{lopes2017exploring}.

\paragraph{Static, analytical, and simulation approaches.}
Analytical GPU models relate execution behavior to program structure and architectural resources \citep{hong2009analytical,baghsorkhi2010adaptive}.
Static PTX analysis supports CUDA execution-time prediction \citep{alavani2018predicting}, while static and predictive analysis guides autotuning without program runs \citep{lim2017autotuning}.
Learned predictors estimate execution time or power from static PTX features \citep{guerreiro2019gpu,tran2025accurate} or instruction statistics collected through instrumented execution \citep{braun2021simple}.
CUDAsap obtains basic-block execution frequencies and instruction statistics statically as an alternative to profiling \citep{emonds2023cudasap}.
TileSight explicitly models tile reuse, cache behavior, and memory traffic \citep{mo2026tilesightfirstprinciplestilecentricanalytical}.
Other approaches use instruction traces for analytical GPU-core modeling, as in GCoM \citep{lee2022gcom}, or detailed GPU simulation, as in PPT-GPU and Accel-Sim \citep{arafa2019pptgpu,khairy2020accelsim}.
Alongside these approaches, we investigate LLMs as estimators of executed arithmetic and device-memory traffic from static evidence.

\paragraph{LLMs for GPU performance prediction.}
Closely related studies by \citet{bolet2025can,bolet2025countingrunningevaluatingllms} evaluate off-the-shelf LLMs for Roofline regime classification and FLOP-count prediction for HeCBench CUDA kernels.
This study extends that setting to estimating device-memory traffic and the numerical \rai derived from FLOPs and traffic.
Trained predictors target related numerical quantities: LLMPerf predicts OpenCL kernel execution time, while Regression Language Models for Code includes Triton kernel latency \citep{nguyennhat2024llmperf,akhauri2026regressionlanguagemodelscode}.
Omniwise fine-tunes a model on HIP kernels to predict arithmetic intensity, bandwidth, cache hit rates, and other performance metrics on AMD MI250 and MI300X GPUs \citep{wang2025omniwise}.
GPU Forecasters evaluates off-the-shelf and reinforcement-learning-tuned LLMs that forecast relative kernel speedup to prioritize GPU measurements during kernel search \citep{khan2026gpuforecasterslanguagemodels}.
We compare source evidence, compiler artifacts, and sandboxed analysis across NVIDIA and AMD targets using off-the-shelf models without task-specific training.
We measure accuracy, cost, latency, and completion without executing the target GPU workload during prediction.

\paragraph{LLMs for parallel code and tool-assisted optimization.}
Complementary work develops high-performance computing code models \citep{nichols2024hpccoder,kadosh2023scope,kadosh2024monocoder}, compiler models for intermediate representations and assembly \citep{cummins2024meta}, and benchmarks for parallel-code and accelerator-kernel generation \citep{nichols2024pareval,ouyang2025kernelbench,wen2025multikernelbench}.
Optimization systems such as CudaForge and AccelOpt use execution or profiling feedback to steer LLM-generated changes \citep{zhang2025cudaforge,nichols2025integrating,zhangAccelOptSelfImprovingLLM}.
CUDA-LLM uses target-GPU latency feedback, while SwizzlePerf supplies hardware specifications and profiling logs \citep{chen2025cudallm,tschand2025swizzleperf}.
AutoKernel profiles models to identify bottlenecks and iteratively tests Triton or CUDA kernel refinements \citep{jaber2026autokernel}.
Kevin trains CUDA-kernel optimization with multi-turn reinforcement learning and execution feedback \citep{baronio2025kevin}.
More broadly, PerfCodeGen refines general code using runtime feedback \citep{peng2025perfcodegen}, and AlphaEvolve combines evolutionary code search with evaluator feedback \citep{novikov2025alphaevolve}.
Mapo trains a memory-performance predictor that uses static features at inference and supplies predicted metrics as rewards to a fine-tuned LLM optimizer \citep{LIU2026108606}.
Such static estimates could complement hardware-feedback optimization by prioritizing candidates while accelerator access is constrained.

\section{Methodology}
\label{sec:methodology}

\begin{figure*}[t]
    \centering
    \includegraphics[width=0.9\linewidth]{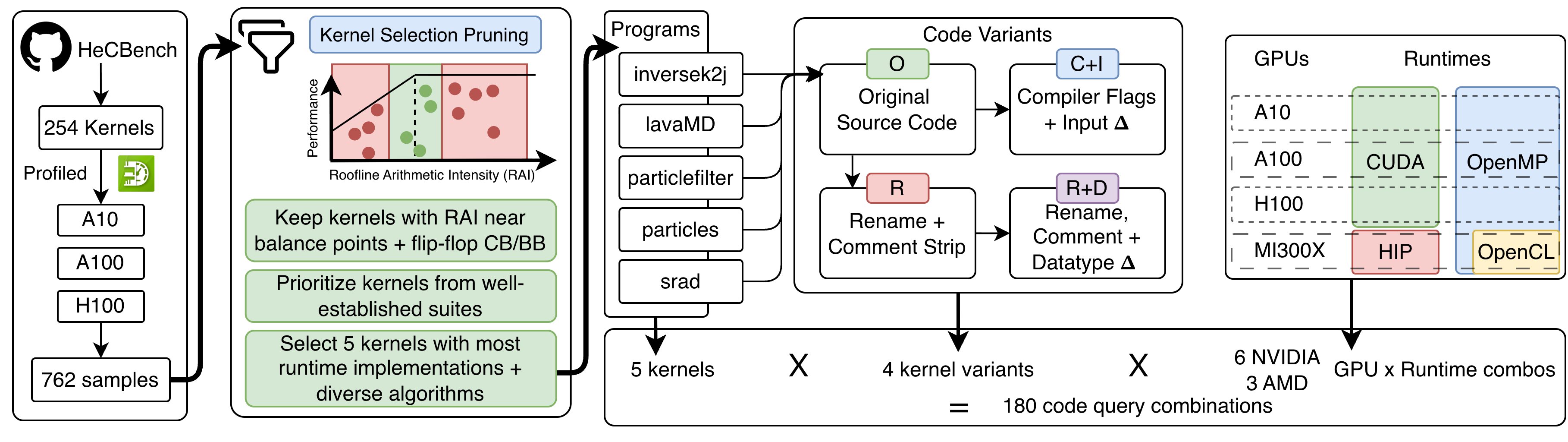}
    \caption{Overview of kernel selection, code variants, and the GPU--runtime combinations used to construct prediction queries.
    Five kernel families, four variants, and nine GPU--runtime combinations define 180 nominal cases; 179 are evaluated because the execution-side OpenCL variant of \texttt{particles} is unavailable.}
    \label{fig:code_combos_diagram}
\end{figure*}

We evaluate how well off-the-shelf LLMs can statically predict GPU kernel Roofline Arithmetic Intensity (\rai) through a series of experiments aimed at:
\textbf{(1)} probing sensitivity to potential memorization of selected evaluation kernels,
\textbf{(2)} testing for any bias towards certain GPU architectures and parallel runtimes, and
\textbf{(3)} avoiding costly LLM experimentation via early pruning of both target GPU kernels and queried LLMs.
These considerations allow us to save both experimental time and money while gathering a meaningful understanding of how off-the-shelf LLMs behave for our target task.
Below we detail the steps and design decisions of our experimentation.

\subsection{Target Kernel Selection}
\label{sec:targetKernelSelection}

\autoref{fig:code_combos_diagram} provides an overview of our kernel selection approach.
Our target kernels are sourced from the \hecbench suite \citep{jin2023hecbench}, where we profiled 254 kernels from well-established suites it contained (e.g.: NAS Parallel Benchmarks, Parboil, Rodinia, NVIDIA CUDA Samples) across three NVIDIA GPUs: an A10, A100, and H100.
\hecbench supplies CUDA, HIP, and OpenMP implementations of the selected programs; OpenCL implementations are drawn from Rodinia or ported for this study, as detailed below.

We used proximity to the GPU balance points and changes in \bbound/\cbound behavior across hardware and parallel runtimes to guide kernel selection.
Cases near a balance point make classification sensitive to numerical \rai errors, while changing regimes require target-specific predictions.
Selection operates at the kernel-family level, so individual variant--GPU--runtime cases can still lie far from their balance points.

Although our initial kernel selection is made with NVIDIA GPUs, we also consider the HIP, OpenMP, and OpenCL variants of the target kernels on an AMD MI300X GPU.
This allows us to check cross-vendor bias in prediction accuracy, given the strong lexical similarity between CUDA/HIP, and GPU-agnostic nature of OpenMP offload kernels.
Unfortunately, OpenCL profiling support had been dropped from the NVIDIA toolchain, thus we sample OpenCL versions of the kernels as an additional \textit{generalizability} study on the MI300X.

Each additional kernel family adds 36 cases, roughly 20\% more queries for every model, strategy, and repetition, so the \$2,000 experimentation budget described in \autoref{sec:llmPruning} led us to limit the study to five families.
Many five-family sets were admissible, so we chose among them with a staged procedure, detailed in \autoref{sec:appendix-kernel-selection}.
We profiled eight candidate families on all four GPUs, each with CUDA, HIP, and OpenMP-offload implementations in \hecbench and an OpenCL implementation that could be sourced or ported.
We excluded families whose dominant-precision FLOP count varied across the NVIDIA GPUs for the same source and input: \texttt{threadfence}, whose contention-dependent atomic retries spread its FLOP count by 37\%, and the integer-only \texttt{bfs}, whose few counted FLOPs also vary by GPU.
We then required each set to contain a saturated compute-bound and a saturated bandwidth-bound anchor, and ranked the four remaining valid sets by the share of their cases within 1.5 octaves ($2.8\times$) of a balance point.
The selected set ranks first (31\% of its cases lie near a balance point), contains no two families with the same work shape and memory-access pattern, and stays first under six of eight alternative ranking choices.

\subsection{Selected GPU Kernels \& Their Variants}

The five selected programs and their target kernels are summarized in \autoref{tab:selected-kernels} in the appendix.
Their upstream lineages span AxBench, Rodinia, and the NVIDIA CUDA Samples: \texttt{inversek2j} comes from AxBench, \texttt{lavaMD}, \texttt{particlefilter}, and \texttt{srad} come from Rodinia, and \texttt{particles} comes from an NVIDIA SDK sample \citep{che2009rodinia,yazdanbakhsh2017axbench,nvidia2018particles}.
The CUDA, HIP, and OpenMP target-offload implementations are taken from \hecbench, while the matched OpenCL implementations use Rodinia donors where available and study-created ports for \texttt{inversek2j} and \texttt{particles}.
Their measured regimes vary with the target GPU, runtime, and datatype.

To test robustness to changes in source code and execution conditions, we study four variants of each program:
(\Ovariant) the \textbf{O}riginal baseline source for each runtime,
(\CIvariant) the same unmodified baseline source with altered \textbf{C}ompiler flags and/or program \textbf{I}nputs, where supported,
(\Rvariant) a comment-stripped version with all symbols/functions \textbf{R}enamed to neutral names,
(\RDvariant) the same comment-stripped/\textbf{R}enamed version with a change in \textbf{D}atatype (from FP32 $\rightarrow$ FP64, and FP64 $\rightarrow$ FP32).
The variants provide complementary comparisons using unchanged and modified source code.
\Ovariant and \CIvariant share identical source, while \Rvariant and \RDvariant remove recognizable names and comments.
\Rvariant preserves the intended computation of \Ovariant, whereas \CIvariant changes compilation settings or inputs and \RDvariant changes the floating-point datatype.
These comparisons test whether predictions respond to surface changes and changes affecting numerical targets, providing a best-effort check of reliance on memorized answers without establishing training-data exposure.

These kernel, variant, GPU, and runtime combinations define 180 nominal cases, of which we evaluate 179.
The OpenCL \texttt{particles} \CIvariant case is excluded because neither its workload dimensions nor its device-compilation options can be changed without modifying the source, which would violate the \CIvariant constraint.

\subsection{LLM Querying Approach}
\label{sec:llmQueryingApproach}

Our evaluation considers \textbf{(1)} direct zero-shot (\textbf{\zeroshot}), and \textbf{(2)} agentic (\textbf{\agentic}) prompting approaches.
An agentic setting is a typical area where we expect our work to be used by bottlenecked systems, often performing agentic kernel optimization in a loop.
We intentionally study \zeroshot alongside \agentic prompting to contrast the cost/time tradeoffs between the two.

\begin{figure}[t]
    \centering
    \includegraphics[width=0.75\linewidth]{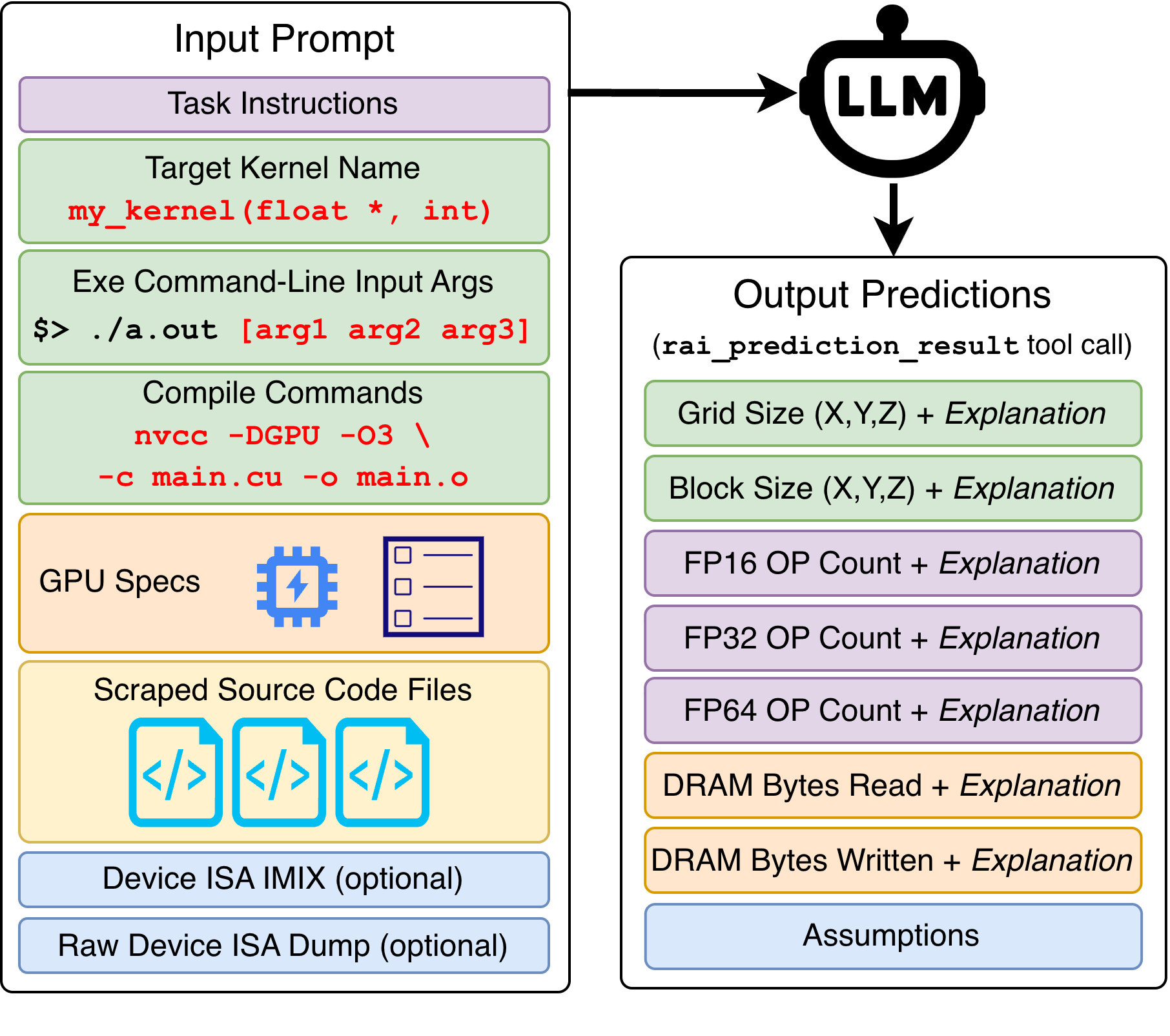}
    \caption{Zero-shot prompting and the expected structured tool-call response.
    Source-only queries provide source code, while source-plus-ISA queries also provide assembly instructions and instruction-mix (IMIX) evidence.}
    \label{fig:zero_shot_prompting_diagram}
\end{figure}

\begin{figure}[t]
    \centering
    \includegraphics[width=0.85\linewidth]{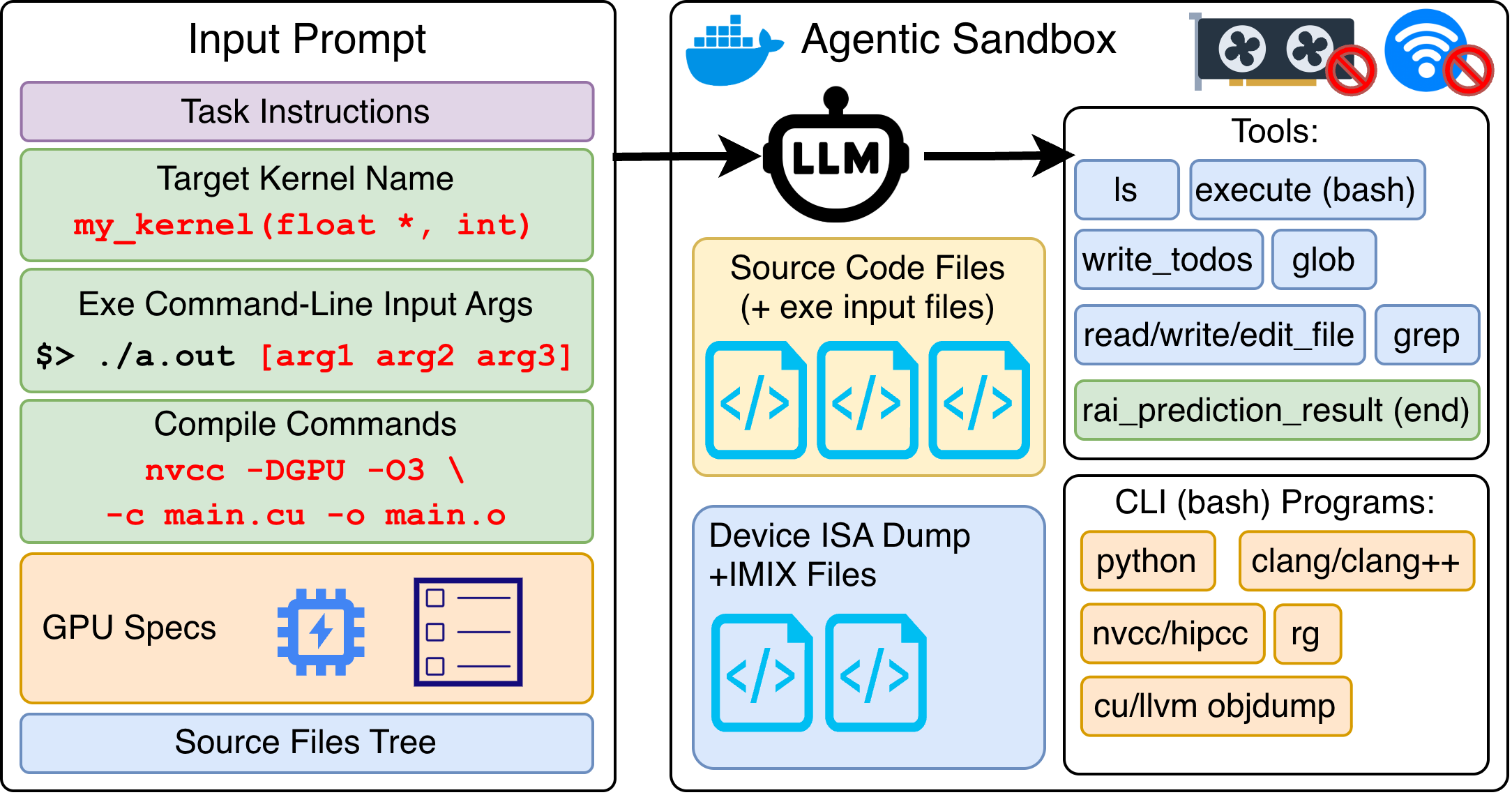}
    \caption{Agentic sandbox setup.
    We keep a similar input prompt as for the \zssc experiments, however we move the verbose source-code and ISA information into the sandbox for the agent to explore.
    Various helper tools are exposed to the LLM for managing a todo-list, reading/writing files, and interacting with the environment's pre-installed CLI programs.
    }
    \label{fig:agent_sandbox_diagram}
\end{figure}

\autoref{fig:zero_shot_prompting_diagram} and \autoref{fig:agent_sandbox_diagram} describe the \zeroshot and \agentic prompting approaches, respectively.
For both cases, we use similar input/instruction prompts to guide the LLMs towards an answer. 
We supply the name of the target kernel, the command line arguments, the compilation commands, and GPU specifications (i.e.: model name, cache hierarchy) so that the LLMs have context w.r.t the kernel we are asking it to predict and the system it is executing on.
\zeroshot queries receive fixed evidence inline, whereas \agentic runs can interactively inspect files and perform computations in the sandbox.
\zeroshot prompts include every source file of the target program, much of which does not affect the target kernel; \agentic runs receive the same files in a sandbox, together with a file tree, and choose what to inspect.
The sandbox has no network access and no GPU, so agents can compile code and run their own analysis scripts but cannot run the target workload on a GPU; each run is limited to 100 model responses.

We additionally study the inclusion of \textit{post-compilation} information in the form of SASS (NVIDIA) and GCN (AMD) instructions corresponding to the target kernels (henceforth referred to as \textbf{ISA}).
Included with the raw ISA dumps, we add static instruction mix (\textbf{IMIX}) information to the \zeroshot prompts and static IMIX files for \agentic.
This IMIX inclusion saves LLMs from having to compute exact static ISA instruction counts, allowing them to focus on reasoning about dynamic execution.
We denote \zeroshot experiments supplying only source-code evidence as \textbf{\zssc} and those supplying ISA as \textbf{\zsisa}.

The expected output from the LLMs is a tool call dubbed \textbf{\toolcall} (shown in \autoref{fig:zero_shot_prompting_diagram}), which contains all the \rai-relevant predictions made by an LLM.
This tool call includes \textbf{(1)} the launch bounds of the target kernel (i.e.: grid/block sizes) to confirm proper thread count estimates, \textbf{(2)} FP16/32/64 FLOP count estimates for the \rai numerator, and \textbf{(3)} bytes read/written across the specified device-memory counter boundary for the \rai denominator.
For post-mortem analysis, each prediction is expected to provide an explanation and a list of assumptions made in an LLM's calculations.
These tool calls normalize our expected outputs, and allow us to compare the ground-truth profiled \rai performance counters to the LLM \rai estimates.

NVIDIA reference counters measure transfers between L2 and DRAM during the first target-kernel dispatch; MI300X counters measure requests from TCC/L2 toward MALL, including requests served by MALL.
The prompts specify these boundaries, so the target is traffic crossing the stated boundary during the dispatch, rather than the byte volume of ISA loads and stores.
A kernel can therefore execute stores yet record zero writes if no write traffic crosses that boundary within the measured interval.

\subsection{Preliminary LLM Pruning Phase}
\label{sec:llmPruning}

\begin{figure}[t]
    \centering
    \begin{subfigure}[b]{0.25\linewidth}
        \centering
        \includegraphics[width=\linewidth]{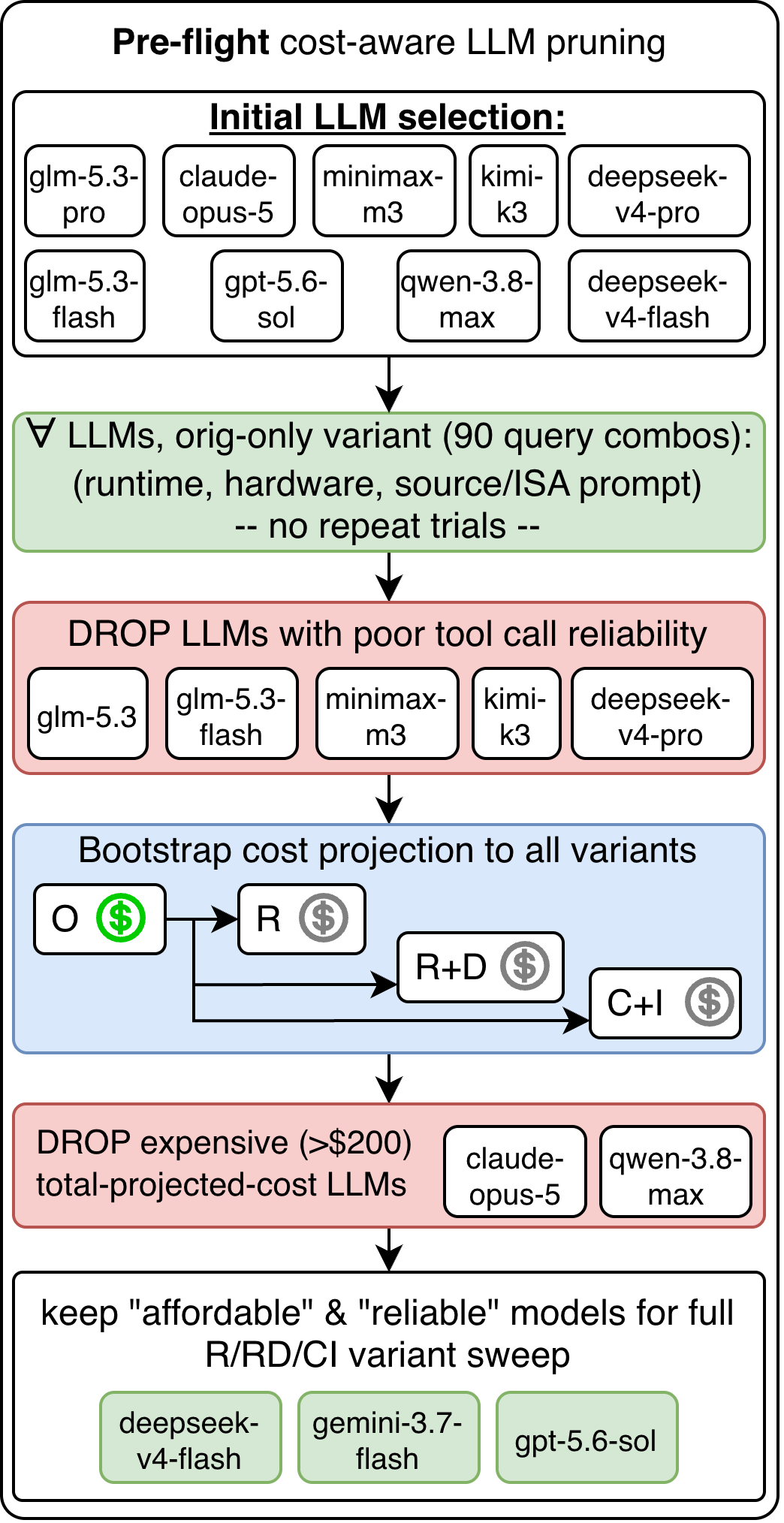}
        \caption{}
        \label{fig:cost_aware_experiment_workflow}
    \end{subfigure}
    \hfill
    \begin{subfigure}[b]{0.73\linewidth}
        \centering
        \includegraphics[width=\linewidth]{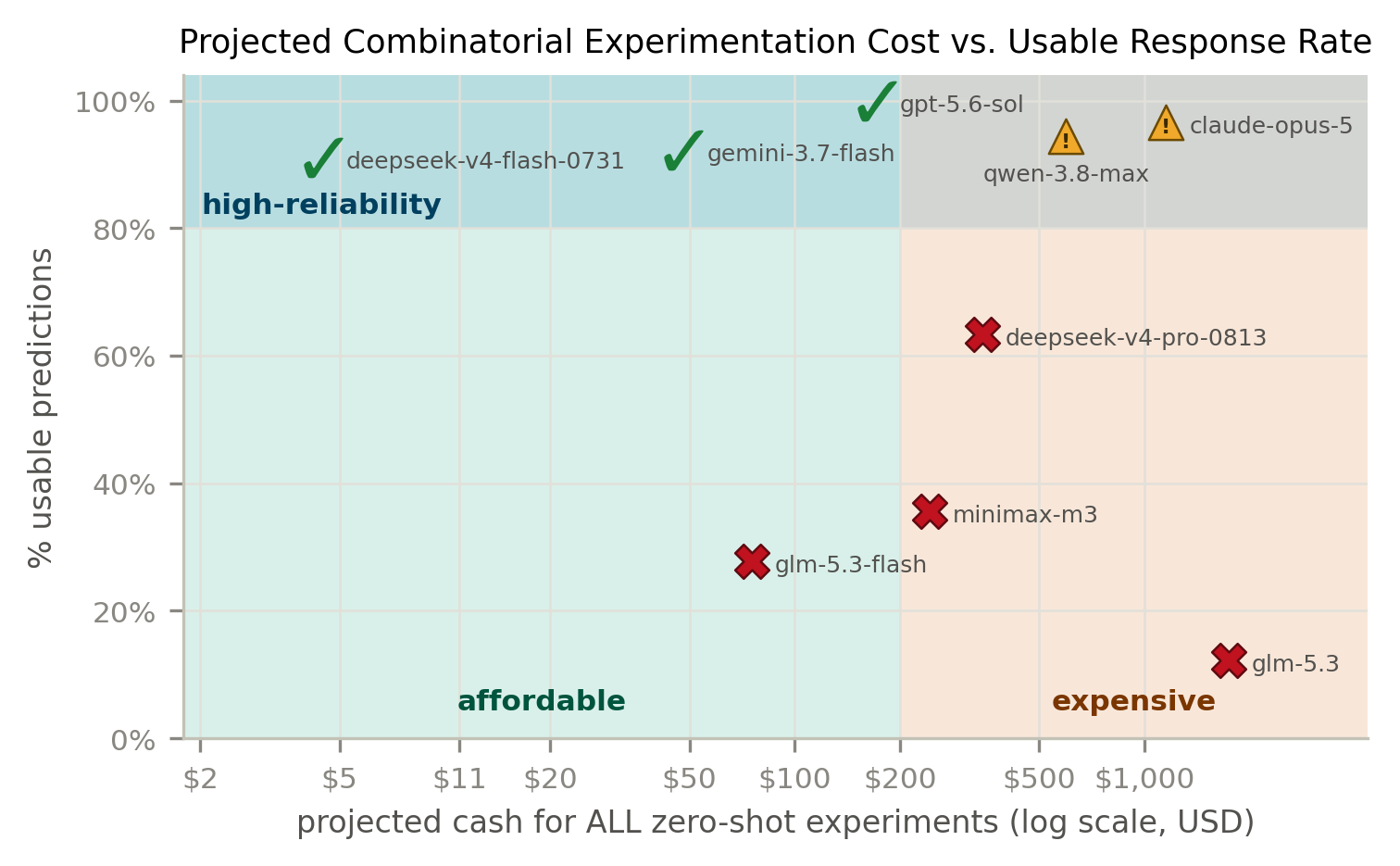}
        \caption{}
        \label{fig:cost_reliability_tradeoff}
    \end{subfigure}
    \caption{(a) pre-flight cost-aware LLM pruning workflow; (b) projected full-sweep cost versus tool-call completion rate.}
    \label{fig:cost_aware_pruning}
\end{figure}

Having 179 evaluated cases and 3 different prompting strategies (\zssc, \zsisa, \agentic) that we want to test across multiple LLMs quickly becomes very expensive, especially when we want to perform repeat trials of each query to account for variation in predictions.
We consider open-source and frontier LLMs with a publication cutoff date of August 18, 2026. 
Appendix \autoref{tab:model-endpoint-comparison} lists the LLMs we initially considered for this work.
For reproducibility purposes, we capped our total experimentation spend at \$2000 USD, with the expectation that the sum of all zero-shot prompting runs for a single LLM should be capped at a reasonable \$200 USD, where leftover total budget would go towards agentic experimentation.
With this spend cap, we pruned costly and unreliable LLMs as shown in \autoref{fig:cost_aware_experiment_workflow}.
We sampled each LLM using only the \Ovariant code variants and both \zssc and \zsisa prompts (a total of 90 query combinations), recording the number of \textit{successfully completed} \toolcall tool calls (i.e., \textit{usable predictions}) and monetary spend for each query.
Each LLM was queried using it's manufacturer-default \textit{thinking} (reasoning) setting, as our experimentation is focused on \textit{out-of-the-box} behavior of these LLMs. 
From this initial cost data, we bootstrap-projected the cost of querying each LLM with the other variants to get a sense of how much the zero-shot experiments would cost us for each model with 5 repetitions of each query; dropping \texttt{claude-opus-5} and \texttt{qwen-3.8-max} as they were projected to cost us \$1,151 and \$597, respectively.
We then dropped the LLMs with less than 80\% usable predictions (\texttt{glm-5.3}, \texttt{glm-5.3-flash}, \texttt{minimax-m3}, \texttt{deepseek-v4-pro-0813}) as they would waste money on incomplete queries.
\autoref{fig:cost_reliability_tradeoff} visualizes the expected spend and percent of usable predictions achieved during this cost-aware LLM pruning phase.

After these pruning steps, we were left with three \textit{reliable} and \textit{cost-effective} LLMs (\deepseekflash, \geminiflash, and \gptsol) that we use for the remainder of our experiments.
We refer to these models as \deepseekshort, \geminishort, and \gptshort, respectively, in the evaluation.

\section{Evaluation}
\label{sec:t-evaluation}
\label{sec:evaluation}

We evaluate 179 kernel--variant--GPU--runtime cases under three models and three strategies, with five repetitions per case: 895 runs per model--strategy configuration and 8,055 runs overall.
A \textit{\textbf{run}} is one prediction attempt, including failed and incomplete predictions.
We compute \rai from total FP16, FP32, and FP64 FLOPs divided by total device-memory bytes read and written.
Traffic counts bytes crossing the target profiler's memory boundary during the kernel dispatch.
Reference ground-truth hardware performance counter values are means taken over ten profiling trials per case; \textbf{\textit{reference}} \rai is computed from these counter means.
For a profiled reference value $x$ (\rai, FLOPs, or bytes) and an LLM estimate $\hat{x}$, the $f\times$ \textbf{\textit{tolerance}} requires $x/f \leq \hat{x} \leq fx$, treating proportional overestimates and underestimates symmetrically.
We choose $1.20\times$ as a primary reporting threshold for close numerical agreement and additionally report a coarser $2\times$ tolerance.
Numerical accuracy is the share of runs meeting the tolerance, with failed or incomplete predictions counted as misses.
Appendix Figure~\ref{fig:t-tolerance} shows that compiler-evidence gains and \gptshort's agentic gain persist at $1.10\times$ and $1.50\times$.

We also evaluate the Roofline classification regime implied by each numerical estimate.
The \textbf{\textit{balance point}} $B$ is the GPU's compute peak for the precision with the largest reference FLOP count divided by its memory-bandwidth peak.
A \rai value below $B$ is bandwidth-bound (\bbound); a \rai value at or above $B$ is compute-bound (\cbound).
Selecting $B$ from each prediction's dominant precision leaves classifications unchanged on the evaluated runs.
Runs without a usable prediction or with non-finite \rai are unclassified and count as classification misses.

RQ1 assesses numerical accuracy, RQ2 the resulting Roofline classifications, RQ3 variation across workloads, GPUs, and code variants, and RQ4 the accuracy, cost, and latency tradeoffs of sandboxed analysis.

\subsection{RQ1: Can LLMs Predict the RAI of GPU Kernels from Static Evidence?}
\label{sec:t-rq1}

\begin{figure}[t]
    \centering
    \includegraphics[width=\linewidth]{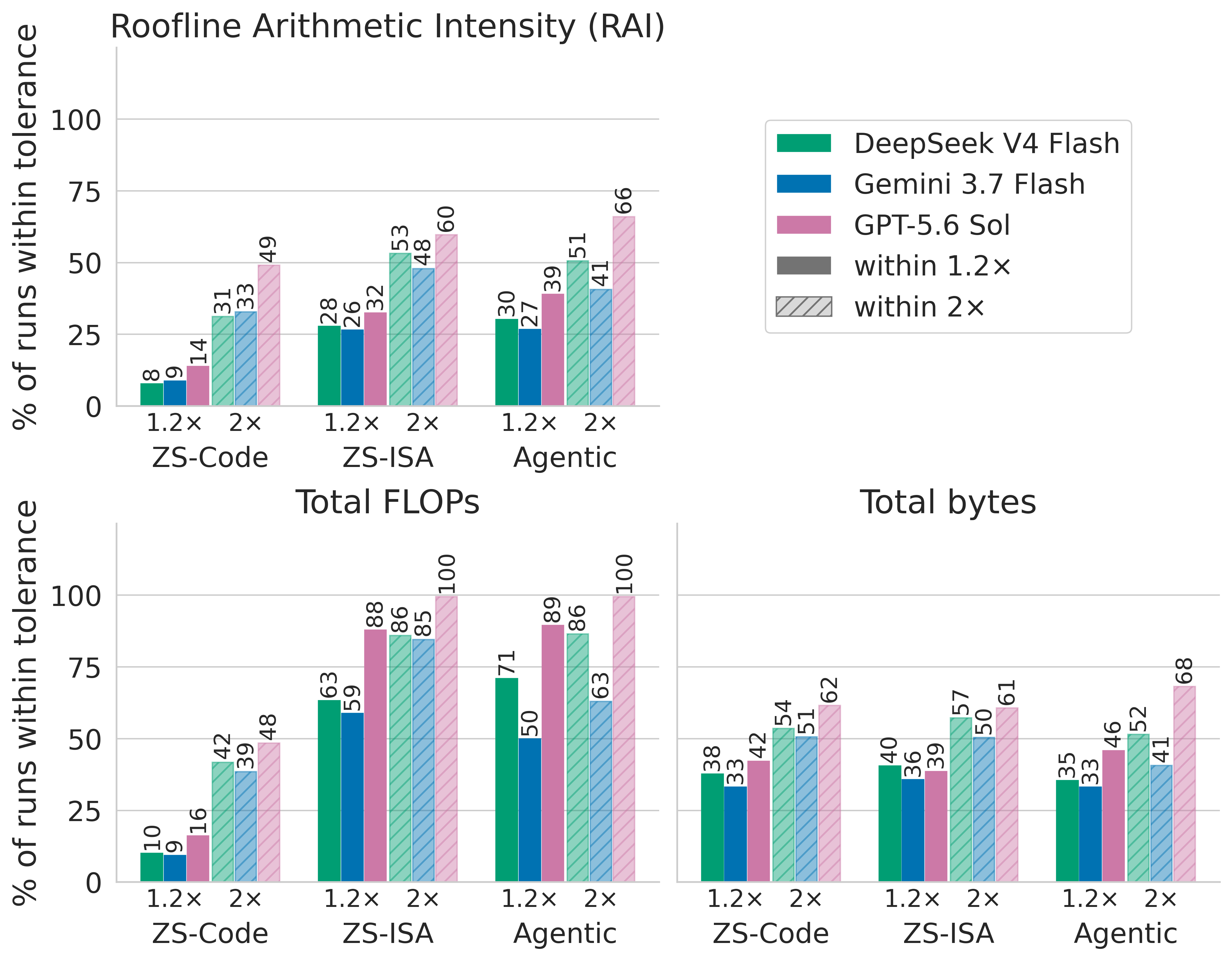}
    \caption{Compiler evidence improves numerical \rai prediction, with larger gains in FLOP counts than in device-memory traffic.
    Panels show \rai (top left), total FLOPs (bottom left; FP16+FP32+FP64), and total device-memory traffic (bottom right; bytes read and written) across models and strategies.
    Solid bars show the percentage of runs within the $1.20\times$ tolerance; hatched bars use the $2\times$ tolerance.
    Failed and incomplete predictions count as misses.
    Appendix \autoref{fig:t-counter_accuracy} breaks FLOPs down by precision and traffic into reads and writes.}
    \label{fig:t-total_accuracy}
\end{figure}

\textbf{With compiler evidence, LLMs predict numerical \rai within $1.20\times$ of the reference in 26--32\% of runs.}
\autoref{fig:t-total_accuracy} compares numerical accuracy for \rai and its FLOP and traffic components across the three models and strategies.
Compiler evidence increases numerical \rai accuracy by 18--20 percentage points over the 8--14\% achieved under \zssc across the three models.
At the $2\times$ tolerance, \zsisa accuracy reaches 48--60\%.
Appendix \autoref{tab:t-accuracy_differences} shows positive 95\% intervals for all three compiler-evidence gains at $1.20\times$ using paired case bootstraps, which resample whole cases with their five runs under both strategies kept together.

\textbf{FLOP estimates benefit more from compiler evidence than memory-traffic estimates.}
Total-FLOP accuracy reaches 59--88\% under \zsisa, with \gptshort improving from 16\% to 88\%.
Under \zsisa, total device-memory traffic is less accurate, at 36--40\%, and remains a recurring source of numerical \rai error.
Unlike an ISA load/store byte total, measured traffic depends on effective cache reuse, eviction, and writeback during the dispatch.

The original-CUDA \texttt{lavaMD} cases illustrate this difficulty: measured traffic is 290\,MB on A10, compared with 149--152\,MB on A100/H100.
For \gptshort under \zsisa, median predicted traffic is $0.92$--$0.93\times$ the reference on A100/H100, within the $1.20\times$ tolerance, but $0.67\times$ on A10, outside the band; Appendix \autoref{tab:t-cases}(a) gives the matched measurements.

\begin{rqsummary}{RQ1}
\textit{Compiler evidence substantially improves numerical \rai prediction, raising agreement within $1.20\times$ by 18--20 percentage points across the tested models.
Under \zsisa, total-FLOP accuracy is 59--88\%, while device-memory traffic accuracy is 36--40\%, leaving traffic estimation a recurring weak point.}
\end{rqsummary}

\subsection{RQ2: Can Predicted RAI Correctly Identify a GPU Kernel's Roofline Regime?}
\label{sec:t-rq2}

\begin{figure}[t]
    \centering
    \includegraphics[width=0.9\linewidth]{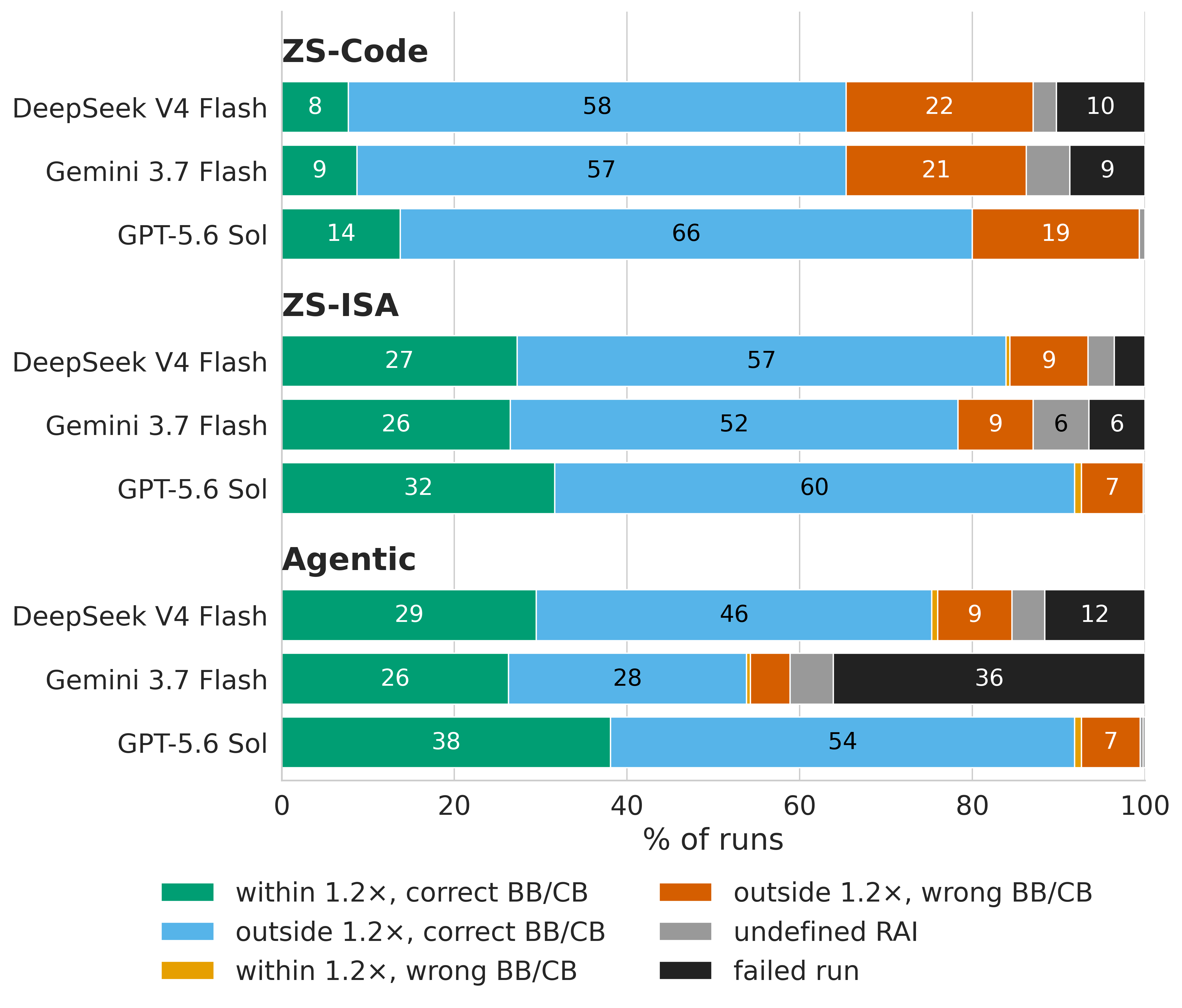}
    \caption{Correct Roofline classifications often survive numerical \rai error.
    Segments separate numerical accuracy at $1.20\times$ from \bbound/\cbound classification outcomes across models and strategies.
    Green and blue segments indicate correct classifications with estimates within and outside the tolerance, respectively; failed runs and undefined \rai estimates are shown separately.
    Segment labels are rounded to whole percentages.}
    \label{fig:t-classification}
\end{figure}

\textbf{Predicted \rai often identifies the correct Roofline regime despite numerical error.}
\autoref{fig:t-classification} separates numerical \rai agreement from \bbound/\cbound classification outcomes for each model and strategy.
For \gptshort under \zsisa, correct classifications inside and outside the tolerance (green plus blue) sum to 91.8\% of runs, compared with 32.4\% whose \rai falls within $1.20\times$ of the reference.
\textbf{In 60.2\% of all \gptshort \zsisa runs, the numerical estimate falls outside this tolerance but the Roofline classification remains correct.}

Classification accuracy rises from 65--80\% under \zssc to 78--92\% under \zsisa across the three models.
Of the 179 evaluated cases, 93 are \bbound and 86 are \cbound by profiled \rai, so always predicting \bbound scores 52\%.
A simple baseline chooses the more common reference class for each GPU--precision combination and reaches 73.2\% accuracy without reading code; Appendix \autoref{tab:t-classification_baselines} gives the group counts.
The eight class choices are fitted and evaluated on the same 179 cases, making this an \textit{in-sample} comparison.
All three models under \zsisa classify individual kernels more accurately than this coarse rule, despite its being fitted to the evaluation labels; under \zssc, only \gptshort does.
Appendix \autoref{tab:t-classification_metrics} confirms that \gptshort's \zsisa accuracy holds for both regimes, with similar recall for \bbound and \cbound.

\begin{figure}[t]
    \centering
    \includegraphics[width=0.7\linewidth]{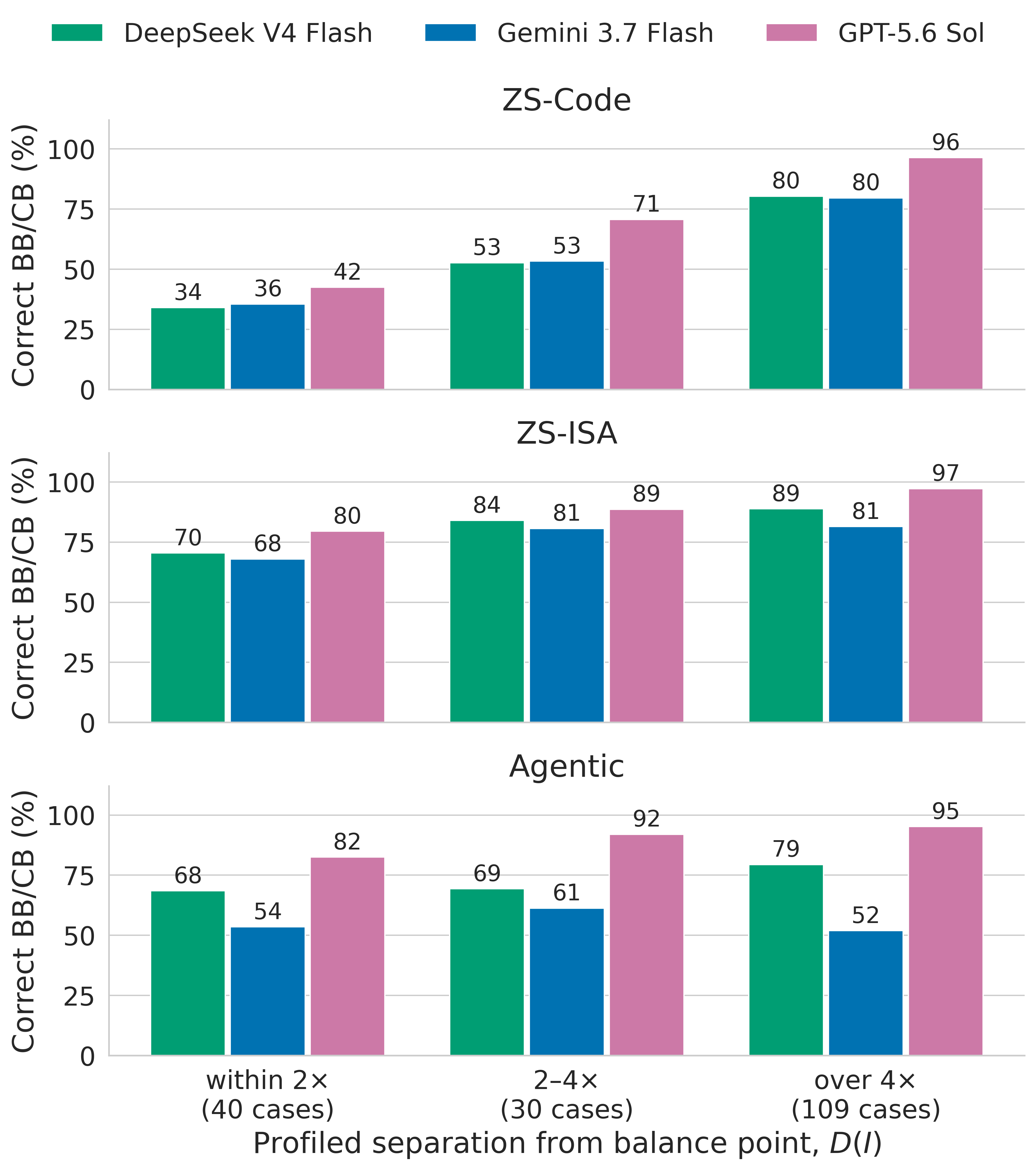}
    \caption{Compiler evidence improves Roofline classification near the balance point.
    Panels compare classification accuracy under \zssc, \zsisa, and \agentic, grouping cases by profiled separation $D(I)\leq2$, $2<D(I)\leq4$, and $D(I)>4$.
    The groups contain 40, 30, and 109 cases, respectively, with five runs per case for each model and strategy.
    Colors distinguish models; bar labels are rounded to whole percentages, and failed or incomplete predictions count as misses.}
    \label{fig:t-ridge_distance}
\end{figure}

\textbf{Compiler evidence helps near the balance point.}
Classification is more sensitive to numerical error near the balance point $B$: a small error can change the predicted regime, whereas farther away a larger error can leave it unchanged.
For a positive \rai $I$, we quantify multiplicative distance from $B$ as $D(I)=\max(I/B,B/I)$.
This factor is symmetric: it equals 1 at the balance point and 4 at $B/4$ or $4B$.
\autoref{fig:t-ridge_distance} groups classification results by their profiled distance $D(I)$, rather than by numerical tolerance.
For cases within $2\times$ of the balance point, \gptshort's classification accuracy rises from 42.5\% under \zssc to 79.5\% under \zsisa, a gain of 37 percentage points.
The paired case-bootstrap 95\% interval for this gain, $[19.5,54.0]$ points, excludes zero, supporting a positive improvement under case resampling.
For cases more than $4\times$ away, \gptshort exceeds 95\% accuracy under every strategy.

\begin{rqsummary}{RQ2}
\textit{Compiler evidence raises \gptshort's classification accuracy from 42.5\% to 79.5\% for the 40 cases within $2\times$ of the balance point.
Across all cases, \zsisa classifications are correct in 78--92\% of runs, often preserving the regime despite numerical error.}
\end{rqsummary}

\subsection{RQ3: Does Numerical RAI Accuracy Vary Across Workloads, Target GPUs, and Code Variants?}
\label{sec:t-rq3}

\begin{figure}[t]
    \centering
    \includegraphics[width=\linewidth]{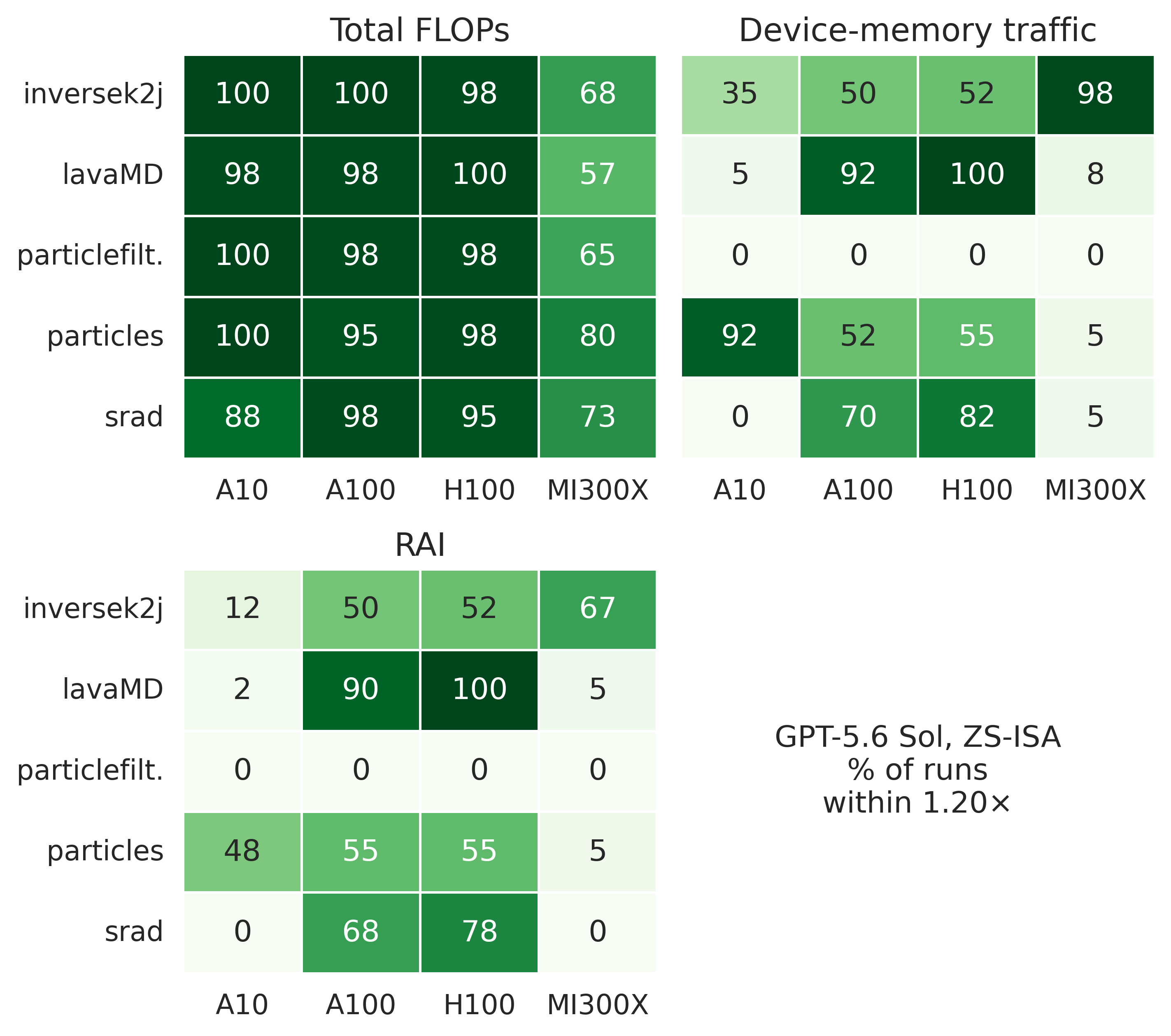}
    \caption{\gptshort's numerical accuracy varies across workload--GPU combinations.
    Panels show \zsisa accuracy within $1.20\times$ of the profiled total FLOPs, total device-memory traffic, and \rai.
    Variants and runtimes are pooled; each NVIDIA cell has 40 runs, each MI300X cell 60, and \texttt{particles} on MI300X 55.
    Failed predictions count as misses.
    Cell values are rounded to whole percentages.
    Runtime sets differ by vendor (CUDA and OpenMP on NVIDIA; HIP, OpenMP, and OpenCL on MI300X), so cells compare the evaluated targets.}
    \label{fig:t-rq3_kernel_gpu}
\end{figure}

\textbf{Numerical \rai accuracy varies across workloads, target GPUs, and code variants.}
\autoref{fig:t-rq3_kernel_gpu} maps \zsisa accuracy for \rai, FLOPs, and traffic across workloads and GPUs for \gptshort.
We focus on the model with the highest overall numerical accuracy to examine its successful cases and remaining limitations.
\texttt{lavaMD} reaches 90\% accuracy within $1.20\times$ on A100 and 100\% on H100, compared with 2.5\% on A10 and 5\% on MI300X.
In contrast, \texttt{particlefilter} has no accurate \rai estimates on any GPU.
The ordering of GPUs also depends on the workload: \texttt{inversek2j} is most accurate on MI300X.
Appendix \autoref{fig:t-joint_accuracy} extends the breakdown to all models and individual runtime--variant combinations; exact GPU rankings differ across models.

\textbf{NVIDIA cells reach 88--100\% total-FLOP accuracy; traffic remains a frequent limitation.}
On MI300X, FLOP-counting errors also contribute: for \texttt{inversek2j}, traffic accuracy reaches 98\%, but FLOP accuracy is 68\% and \rai accuracy is 67\%.
FLOP accuracy also requires matching the target profiler's counting convention.
Appendix \autoref{tab:inversek2j-accounting} illustrates this in checked \gptshort \zsisa \texttt{inversek2j} predictions: mismatched accounting for packed arithmetic and division helpers causes both FLOP and \rai estimates to miss tolerance.
Component errors can compound: estimates within $1.20\times$ for both FLOPs and traffic guarantee only a $1.44\times$ bound on \rai.
For \gptshort on \texttt{particles}/A10, 18 of 40 \zsisa runs meet both component tolerances yet miss the \rai target.

Repeated accesses to neighboring boxes offer cache reuse in \texttt{lavaMD}.
The \Ovariant CUDA \texttt{particlefilter} writes and rereads a 110\,MB temporary array, with adjacent threads accessing entries 276 bytes apart.
These strided accesses make traffic sensitive to transaction granularity and effective reuse, even when FLOP estimates are accurate.
Appendix \autoref{tab:t-cases}(a) shows median traffic predictions 4--26$\times$ below the reference for the \Ovariant CUDA/HIP \texttt{particlefilter} cases across all four GPUs.

\begin{figure}[t]
    \centering
    \includegraphics[width=0.8\linewidth]{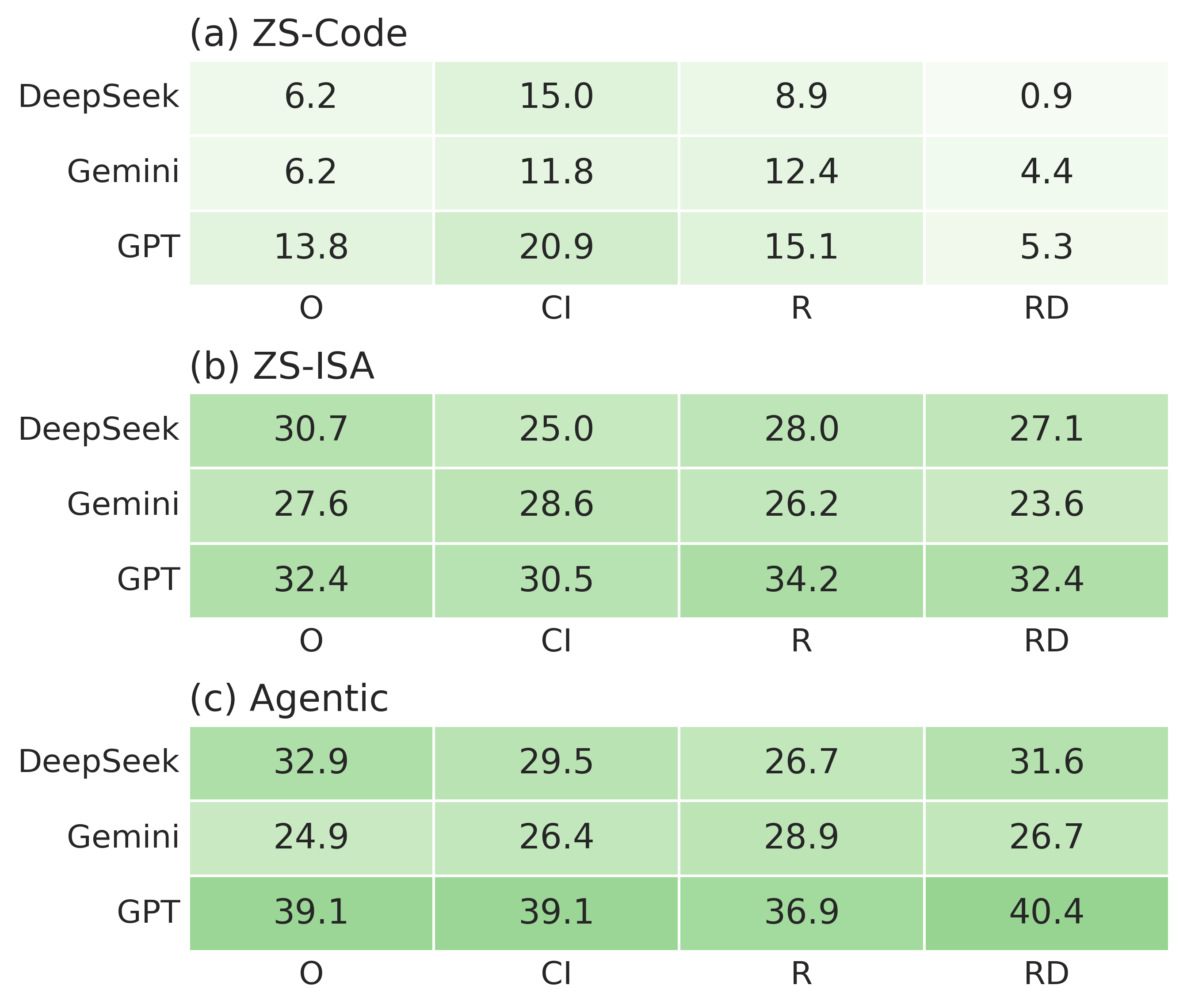}
    \caption{Compiler-evidence gains extend across all four code variants.
    Cells show numerical \rai accuracy within $1.20\times$ (\%) for each model and strategy, pooling workloads, GPUs, and runtimes.
    Columns show original (\Ovariant), altered flags/inputs (\CIvariant), renamed (\Rvariant), and renamed with changed datatype (\RDvariant) variants.
    Each cell includes 225 runs, except \CIvariant with 220; failed and incomplete predictions count as misses.}
    \label{fig:t-variant_accuracy}
\end{figure}

\textbf{Compiler evidence improves accuracy in all variants.}
\autoref{fig:t-variant_accuracy} compares \rai accuracy for every model and strategy across variants that alter names, compiler flags, inputs, or datatypes.
\zsisa exceeds \zssc for every model--variant combination.
\CIvariant improves \zssc accuracy relative to \Ovariant for all three models but has mixed effects under \zsisa.
Relative to \Ovariant, \Rvariant has no consistent directional effect on accuracy across models and strategies.
\RDvariant reduces \zssc accuracy relative to \Ovariant, most strongly for \deepseekshort and \gptshort; \gptshort's \Ovariant and \RDvariant accuracies rise from 13.8\% and 5.3\% under \zssc, respectively, to 32.4\% for both under \zsisa.
Appendix \autoref{tab:t-variants} shows near-complete predictions across \gptshort's variants but variant-dependent completion for the other models.

\begin{rqsummary}{RQ3}
\textit{Numerical \rai accuracy varies across workloads and targets: \gptshort under \zsisa reaches 90--100\% accuracy within $1.20\times$ for \texttt{lavaMD} on A100/H100, but 0\% for \texttt{particlefilter} on every GPU.
Compiler-evidence gains extend across all four code variants, while traffic estimation remains a recurring weak point.}
\end{rqsummary}

\subsection{RQ4: Can Sandboxed Analysis Improve Numerical RAI Accuracy Beyond Zero-Shot ISA Prompting?}
\label{sec:t-rq4}

\newcommand{\agenticCostRatioMin}{3.6}
\newcommand{\agenticCostRatioMax}{17.7}
\newcommand{\deepseekAgenticCostRatio}{6.7}
\newcommand{\deepseekAgenticLatencyRatio}{2.9}
\newcommand{\geminiAgenticCostRatio}{17.7}
\newcommand{\geminiAgenticLatencyRatio}{12.1}
\newcommand{\gptAgenticCostRatio}{3.6}
\newcommand{\gptAgenticLatencyRatio}{2.5}

\begin{table}[t]
    \caption{Accuracy, completion, recorded cost, and prediction latency under \zsisa and \agentic (895 runs per configuration).
    Cost and latency are medians, with ratios calculated before rounding; \rai accuracy uses the $1.20\times$ tolerance.
    Completion denotes a usable prediction.
    Appendix Figures~\ref{fig:recorded_cost_distribution} and~\ref{fig:prediction_latency} show the corresponding per-run distributions.
    Missing charges are omitted.}
    \label{tab:t-tradeoffs}
    \centering
    \begingroup
\footnotesize
\setlength{\tabcolsep}{3pt}
\begin{tabular}{@{}lrrrr@{}}
Strategy & \shortstack{Cost/run (USD)\\Median} & \shortstack{Latency (min)\\Median} & \shortstack{RAI acc.\\(\%)} & \shortstack{Complete\\(\%)} \\
\midrule
\multicolumn{5}{@{}l}{\deepseekshort} \\
\zsisa & $0.0060$ & $7.6$ & 27.7 & 96.4 \\
\agentic & $0.0405$ & $22.0$ & 30.2 & 88.4 \\
\midrule
\multicolumn{5}{@{}l}{\geminishort} \\
\zsisa & $0.0400$ & $0.62$ & 26.5 & 93.5 \\
\agentic & $0.709$ & $7.6$ & 26.7 & 63.9 \\
\midrule
\multicolumn{5}{@{}l}{\gptshort} \\
\zsisa & $0.223$ & $1.3$ & 32.4 & 100.0 \\
\agentic & $0.800$ & $3.4$ & 38.9 & 99.8 \\
\end{tabular}
\endgroup

\end{table}

\textbf{Agentic prompting improves end-to-end numerical \rai accuracy for \gptshort, while gains for the other models remain uncertain.}
\autoref{tab:t-tradeoffs} compares the accuracy, completion, cost, and latency of both strategies for all three models.
For \gptshort, accuracy within $1.20\times$ rises from 32.4\% to 38.9\%, a gain of 6.5 percentage points with a paired case-bootstrap 95\% interval of $[3.4,9.6]$ points.
The improvement comes at approximately $\gptAgenticCostRatio\times$ the median recorded cost and $\gptAgenticLatencyRatio\times$ the median latency of \zsisa.

The observed \agentic \rai accuracy increases are 2.5 points for \deepseekshort and 0.2 for \geminishort.
Appendix~\ref{sec:t-bootstrap} shows that case-bootstrap intervals for these differences span both decreases and increases, leaving their gains uncertain.
\textbf{Lower completion also contributes to their end-to-end results: \agentic completion falls to 88\% and 64\%, respectively.}
Classification accuracy also falls to 75\% for \deepseekshort and 54\% for \geminishort, while \gptshort remains at 91.8\%.
Appendix \autoref{fig:run_outcomes} shows that exhausting the 100-response budget accounts for most unsuccessful \agentic runs for \deepseekshort and \geminishort.

The sandbox enables analysis beyond a fixed prompt: inspected \gptshort \texttt{lavaMD} trajectories construct GPU-specific LRU cache-simulator code.
Appendix~\ref{sec:t-cache-sim} documents differing accuracy associations across GPUs; the observational comparison does not isolate a simulation benefit.

Under \zsisa, \gptshort has the highest numerical \rai accuracy, \deepseekshort the lowest median recorded cost, and \geminishort the lowest median latency.
\textbf{\agentic multiplies median cost per run by $\agenticCostRatioMin$--$\agenticCostRatioMax\times$, with a clearly supported gain of 6.5 percentage points in numerical accuracy only for \gptshort.}

\begin{rqsummary}{RQ4}
\textit{\agentic improves \gptshort's numerical \rai accuracy within $1.20\times$ by 6.5 percentage points over \zsisa at $\gptAgenticCostRatio\times$ the median cost.
For \deepseekshort and \geminishort, \agentic costs $\deepseekAgenticCostRatio$--$\geminiAgenticCostRatio\times$ as much as \zsisa, with uncertain gains and lower completion.}
\end{rqsummary}

\subsection{Scope and Validity}
\label{sec:t-validity}

Within a \$2,000 experimentation budget, we prioritized comparisons across hardware, runtimes, code variants, and strategies by limiting the study to five kernel families and three LLMs.
Repeated predictions characterize variability within these families rather than additional workload diversity.
Of the 179 cases, 109 lie more than $4\times$ from the profiled balance point, so overall classification accuracy should be read alongside the distance breakdown.

Across ten profiling trials per case, \rai's per-case coefficient of variation has a median of 0.08\% and a maximum of 10.3\%; Appendix~\ref{sec:t-reference-variability} gives the GPU breakdown.
We use fixed prompt designs per strategy and do not assess sensitivity to alternative wording.
We do not compare numerical estimates against dedicated static FLOP-counting or memory-traffic models.

\section{Conclusion}
\label{sec:conclusion}

In this work, we show that LLMs can estimate the Roofline Arithmetic Intensity of GPU kernels from static evidence across different hardware targets, parallel runtimes, and modified code variants.
We empirically show that prompting LLMs with post-compilation assembly instruction artifacts substantially improves arithmetic intensity estimates compared to prompting with source code alone.
Although estimates often contain errors, the resulting Roofline bandwidth- or compute-bound classifications remain correct in 78--92\% of runs, depending on the LLM used for prediction.
Our results highlight that hardware-specific floating point operation-counting conventions can cause FLOP estimation errors, while device-memory traffic estimation remains the main limitation of this approach, because it requires accounting for effective cache reuse at the target profiler's counter boundary.
We also emphasize that compared to zero-shot prompting, agentic LLM prompting in some cases can be detrimental to finishing a task, and in others it only provides disproportionately small improvements in arithmetic intensity prediction accuracy while costing up to $\agenticCostRatioMax\times$ as much.
LLMs can thus help identify a kernel's likely Roofline bandwidth- or compute-bound regime from static evidence, while cache-simulation and memory-traffic tools warrant further study as ways to improve arithmetic intensity numerical estimates.

\section*{Acknowledgments}

This research was supported by Code Metal Inc.\ and the Pazy Foundation.
This work was performed in part under the auspices of the U.S. Department of Energy by Lawrence Livermore National Laboratory under Contract DE-AC52-07NA27344 (LLNL-JRNL-2026894).
The views and opinions of the authors do not necessarily reflect those of the U.S. government or Lawrence Livermore National Security, LLC neither of whom nor any of their employees make any endorsements, express or implied warranties or representations or assume any legal liability or responsibility for the accuracy, completeness, or usefulness of the information contained herein.
The United States Government retains, and the publisher, by accepting the article for publication, acknowledges that the United States Government retains a non-exclusive, paid-up, irrevocable, world-wide license to publish or reproduce the published form of this manuscript, or allow others to do so, for United States Government purposes.

\clearpage
\bibliographystyle{mlsys2025}
\bibliography{refs}
\section*{AI Usage Disclosure}

We used generative AI both as the subject of the experiments and as assistance in implementing the study and preparing the manuscript.
The models evaluated as predictors, their inputs, and their role in producing the experimental outputs are described in the methodology.
The following describes AI assistance used to produce the work.
We used OpenAI's Codex assistant and agents using Anthropic's Claude Opus models for implementation and manuscript preparation.

\paragraph{Research planning and implementation.}
Under author direction, LLM agents wrote substantial portions of the experimental implementation and data-collection software.
AI assistants also helped develop analysis scripts, generate figures and tables, and check reported results against analysis outputs.
They proposed refinements to metrics, statistical analyses, visualizations, and the presentation of findings; the authors selected which suggestions to adopt.

\paragraph{Benchmarks and data.}
The reference FLOP and device-memory traffic values are hardware performance-counter measurements collected with vendor profilers.
LLM-produced FLOP and traffic estimates are the predictions evaluated against those references.
Agents also assisted with the study-created OpenCL ports.
The authors checked these ports by comparing their ground-truth performance profiles with those of the original implementations, aiming to match the original performance as closely as possible while preserving source semantics.

\paragraph{Writing and literature search.}
The authors wrote initial drafts of all the paper sections and used AI assistants to revise, edit, and critique the manuscript.
Assistance also included developing outlines, drafting and substantially rewriting selected passages under author guidance, and refining figure and table captions.
The authors iteratively reviewed the suggestions, supplied revisions, and made the final editorial decisions.
AI assistants helped identify overlooked related papers, retrieve bibliographic records, and check citations against available primary sources.

The authors take full responsibility for the correctness, originality, and integrity of the paper, its software, and its reported results, including material produced with AI assistance.

\clearpage
\appendix
\renewcommand{\thetable}{\Alph{section}.\arabic{table}}
\renewcommand{\thefigure}{\Alph{section}.\arabic{figure}}
\section{Additional Details}
\label{sec:appendix}

\subsection{Model and Endpoint Configuration}
\label{sec:appendix-model-config}

Table~\ref{tab:model-endpoint-comparison} records the model and endpoint configuration snapshot used for experiment planning.
Prices are the frozen, uncached standard rates in our study registry, expressed per one million input and output tokens; they are not necessarily the cheapest route currently advertised by a gateway.
Resource comparisons use recorded experiment costs, including charges estimated from reported usage.
Provider catalogs and endpoint limits are volatile, so we report the snapshot date and distinguish model-level limits from route-dependent limits~\citep{openrouterModels2026,openrouterReasoning2026,openaiGPT55Models2026,azureFoundryModels2026}.

\begin{table*}[t]
  \caption{Model and endpoint comparison (configuration snapshot: 2026-08-24).
  Context is generally the combined input--output window, rather than an independently usable input allowance.
  Prices marked \textsuperscript{C} are conservative route-planning rates; \textsuperscript{S} denotes a shadow/list-price valuation paid with project credits rather than study cash.}
  \label{tab:model-endpoint-comparison}
  \centering
  \scriptsize
  \setlength{\tabcolsep}{2.5pt}
  \renewcommand{\arraystretch}{1.16}
  \begin{tabular}{@{}>{\raggedright\arraybackslash}p{0.15\textwidth}>{\raggedright\arraybackslash}p{0.085\textwidth}>{\raggedright\arraybackslash}p{0.13\textwidth}>{\raggedright\arraybackslash}p{0.15\textwidth}>{\raggedright\arraybackslash}p{0.14\textwidth}>{\raggedright\arraybackslash}p{0.265\textwidth}@{}}
    Model & Release & Price / 1M tokens (input/output) & Context limit & Output limit & Reasoning-token limit \\
    \midrule
    GPT-5.6 Sol
      & 2026-07-09
      & \$5.50 / \$33.00\textsuperscript{C}
      & 1,050,000
      & 128,000
      & No separately published limit; reasoning and visible output share the completion budget. \\
    Claude Opus 5
      & 2026-07-24
      & \$5.00 / \$25.00
      & 1,000,000
      & 128,000
      & OpenRouter caps an explicit reasoning allocation at 128,000; the final answer must also fit. \\
    Kimi K3
      & 2026-07-16
      & \$3.00 / \$15.00\textsuperscript{C}
      & 1,048,576
      & Route-dependent; catalog maximum approximately 974,842
      & No separately published limit; constrained by the served route's completion budget. \\
    GPT-5.5 (Azure)
      & 2026-04-24
      & \$5.00 / \$30.00\textsuperscript{S}
      & 1,050,000 model; approximately 922,000 effective Azure input budget
      & 128,000
      & No separate limit; reasoning and visible output share the 128,000-token generation budget. \\
    Gemini 3.7 Flash
      & 2026-08-13
      & \$0.375 / \$1.875
      & 1,048,576
      & 65,536
      & No public per-effort cap; the provider determines reasoning use internally. \\
    Qwen 3.8 Max
      & 2026-08-03
      & \$2.00 / \$6.00
      & 1,000,000
      & 131,072
      & No separately published limit; constrained by the completion budget. \\
    GLM-5.3
      & 2026-08-18
      & \$1.40 / \$4.40
      & 1,048,576
      & 131,072 advertised; 65,536 effective default observed
      & No separately published limit; constrained by the completion budget. \\
    DeepSeek V4 Pro 0813
      & 2026-08-12
      & \$1.32 / \$3.96\textsuperscript{C}
      & 1,048,576
      & 384,000 catalog maximum; route-dependent
      & No separately published limit; constrained by the served route. \\
    DeepSeek V4 Flash 0731
      & 2026-07-31
      & \$0.44 / \$1.32\textsuperscript{C}
      & Up to 1,310,720; route-dependent
      & 262,144 catalog maximum; route-dependent
      & No separately published limit; constrained by the served route. \\
    MiniMax M3
      & 2026-05-31
      & \$0.75 / \$3.00\textsuperscript{C}
      & Up to 1,048,576; route-dependent
      & 262,144 catalog maximum; route-dependent
      & No separately published limit; constrained by the served route. \\
  \end{tabular}
\end{table*}

The absence of a separate reasoning-token limit does not imply unlimited reasoning.
In those rows, hidden reasoning competes with visible output under the endpoint's completion limit.
OpenRouter effort labels may influence a provider's allocation, but they do not establish portable hard token limits.
For route-dependent models, the exact provider route and response-reported usage must therefore accompany empirical cost and latency results.
In particular, although GLM-5.3 advertises a 131,072-token maximum output, our requests without a client-imposed output limit have observed an effective 65,536-token stopping point; we treat that as endpoint behavior, not as a universal model limit.

\subsection{Selected Target Kernels and Hardware}
\label{sec:appendix-selected-kernels}

Table~\ref{tab:selected-kernels} summarizes the five selected kernel families, their implementation provenance, their measured Roofline regimes across variants, and the profiled GPU specifications.

\begin{table*}[t]
    \centering
    \caption{Selected target kernels, implementation provenance, measured Roofline behavior, and the GPUs those measurements were taken on.}
    \label{tab:selected-kernels}

    \begin{subtable}[t]{\linewidth}
    \centering
    \caption{Identity, provenance, and operation.}
    \label{tab:selected-kernel-provenance}
    \footnotesize
    \setlength{\tabcolsep}{3.5pt}
    \renewcommand{\arraystretch}{1.2}
    \newcolumntype{P}[1]{>{\raggedright\arraybackslash}p{#1\linewidth}}
    \begin{tabular}{@{}P{0.20}P{0.225}P{0.10}P{0.115}P{0.095}P{0.16}@{}}
        & & & \multicolumn{2}{c}{\begin{tabular}[b]{@{}c@{}}Implementation\\ source\end{tabular}} & \\
        \cmidrule(lr){4-5}
        Program & Target kernel by API & Upstream suite & CUDA/HIP/\newline OMP & OpenCL & Operation \\
        \midrule
        \addlinespace[2pt]
        \textbf{\texttt{inversek2j}} & CUDA/HIP/OpenCL:\newline \texttt{invkin\_kernel} & AxBench & HeCBench & Study port & Two-joint inverse kinematics \\
        \addlinespace[4pt]
        \textbf{\texttt{lavaMD}} & CUDA/HIP: \texttt{md}\newline OpenCL:\newline \texttt{kernel\_gpu\_opencl} & Rodinia & HeCBench & Rodinia donor & Neighbor-box molecular forces \\
        \addlinespace[4pt]
        \textbf{\texttt{particlefilter}} & CUDA/HIP:\newline \texttt{kernel\_likelihood}\newline OpenCL:\newline \texttt{likelihood\_kernel} & Rodinia & HeCBench & Rodinia donor & Particle likelihood update \\
        \addlinespace[4pt]
        \textbf{\texttt{particles}} & CUDA/HIP/OpenCL:\newline \texttt{collideK} & NVIDIA CUDA Samples & HeCBench & Study port & Particle collision response \\
        \addlinespace[4pt]
        \textbf{\texttt{srad}} & CUDA/HIP: \texttt{srad}\newline OpenCL:\newline \texttt{srad\_kernel} & Rodinia & HeCBench & Rodinia donor & Anisotropic diffusion coefficient \\
        \addlinespace[2pt]
    \end{tabular}
\end{subtable}

    \vspace{1.0em}

    \begin{subtable}[t]{0.47\linewidth}
    \centering
    \caption{Measured bound-class signatures: \CBmark{} compute-bound, \BBmark{} bandwidth-bound, and \NCmark{} not collected.}
    \label{tab:selected-kernel-regimes}
    \footnotesize
    \setlength{\tabcolsep}{2pt}
    \renewcommand{\arraystretch}{1.15}
    \newcommand{\rothead}[1]{\makebox[0pt][r]{\rotatebox[origin=rB]{-45}{\texttt{#1}\,}}}
    \begin{tabular}{@{}llccccc@{}}
        GPU & API
          & \rothead{inversek2j} & \rothead{lavaMD} & \rothead{particlefilter}
          & \rothead{particles} & \rothead{srad} \\
        \midrule
        A10 & CUDA & \regimesig{\BBmark}{\BBmark}{\CBmark}{\BBmark} & \regimesig{\CBmark}{\CBmark}{\CBmark}{\CBmark} & \regimesig{\BBmark}{\BBmark}{\BBmark}{\BBmark} & \regimesig{\BBmark}{\BBmark}{\CBmark}{\BBmark} & \regimesig{\BBmark}{\BBmark}{\CBmark}{\BBmark} \\
            & OMP & \regimesig{\BBmark}{\BBmark}{\CBmark}{\BBmark} & \regimesig{\CBmark}{\CBmark}{\CBmark}{\CBmark} & \regimesig{\BBmark}{\BBmark}{\CBmark}{\BBmark} & \regimesig{\BBmark}{\BBmark}{\CBmark}{\BBmark} & \regimesig{\BBmark}{\BBmark}{\CBmark}{\BBmark} \\
        \addlinespace[2pt]
        A100 & CUDA & \regimesig{\CBmark}{\CBmark}{\CBmark}{\CBmark} & \regimesig{\CBmark}{\CBmark}{\CBmark}{\CBmark} & \regimesig{\BBmark}{\BBmark}{\BBmark}{\BBmark} & \regimesig{\CBmark}{\CBmark}{\CBmark}{\CBmark} & \regimesig{\CBmark}{\CBmark}{\CBmark}{\BBmark} \\
             & OMP & \regimesig{\CBmark}{\CBmark}{\CBmark}{\CBmark} & \regimesig{\CBmark}{\CBmark}{\CBmark}{\CBmark} & \regimesig{\BBmark}{\BBmark}{\BBmark}{\BBmark} & \regimesig{\CBmark}{\CBmark}{\CBmark}{\CBmark} & \regimesig{\CBmark}{\CBmark}{\CBmark}{\CBmark} \\
        \addlinespace[2pt]
        H100 & CUDA & \regimesig{\CBmark}{\CBmark}{\CBmark}{\BBmark} & \regimesig{\CBmark}{\CBmark}{\CBmark}{\CBmark} & \regimesig{\BBmark}{\BBmark}{\BBmark}{\BBmark} & \regimesig{\CBmark}{\CBmark}{\CBmark}{\BBmark} & \regimesig{\CBmark}{\CBmark}{\CBmark}{\BBmark} \\
             & OMP & \regimesig{\CBmark}{\CBmark}{\CBmark}{\BBmark} & \regimesig{\CBmark}{\CBmark}{\CBmark}{\CBmark} & \regimesig{\BBmark}{\BBmark}{\BBmark}{\BBmark} & \regimesig{\BBmark}{\BBmark}{\CBmark}{\BBmark} & \regimesig{\CBmark}{\CBmark}{\CBmark}{\BBmark} \\
        \addlinespace[2pt]
        MI300X & OMP & \regimesig{\BBmark}{\BBmark}{\BBmark}{\BBmark} & \regimesig{\CBmark}{\CBmark}{\CBmark}{\CBmark} & \regimesig{\BBmark}{\BBmark}{\BBmark}{\BBmark} & \regimesig{\BBmark}{\BBmark}{\BBmark}{\BBmark} & \regimesig{\BBmark}{\BBmark}{\BBmark}{\BBmark} \\
               & HIP & \regimesig{\BBmark}{\BBmark}{\BBmark}{\BBmark} & \regimesig{\CBmark}{\CBmark}{\CBmark}{\CBmark} & \regimesig{\BBmark}{\BBmark}{\BBmark}{\BBmark} & \regimesig{\BBmark}{\BBmark}{\BBmark}{\BBmark} & \regimesig{\BBmark}{\BBmark}{\BBmark}{\BBmark} \\
               & OpenCL & \regimesig{\CBmark}{\CBmark}{\CBmark}{\CBmark} & \regimesig{\CBmark}{\CBmark}{\CBmark}{\CBmark} & \regimesig{\BBmark}{\BBmark}{\BBmark}{\BBmark} & \regimesig{\BBmark}{\BBmark}{\BBmark}{\NCmark} & \regimesig{\BBmark}{\BBmark}{\BBmark}{\BBmark} \\
    \end{tabular}
    \par\vspace{0.35em}\raggedright\footnotesize Within-cell order is original, rename-only (R), rename plus datatype (RD), and compiler plus input (CI). Each arm is classified against the target GPU's ridge point at that arm's dominant precision.
\end{subtable}

    \hfill
    \begin{subtable}[t]{0.51\linewidth}
    \centering
    \caption{Profiled GPUs and the DRAM-level balance points used for BB/CB classification. Peak throughput is in TFLOP/s; balance points (BP) are in FLOP/byte.}
    \label{tab:selected-kernel-gpus}
    \footnotesize
    \setlength{\tabcolsep}{2.5pt}
    \renewcommand{\arraystretch}{1.15}
    \begin{tabular}{@{}lrrrr@{}}
        & A10 & A100 & H100 & MI300X \\
        \midrule
        Architecture & GA102 & GA100 & GH100 & CDNA3 \\
        Compute cap. & 8.6 & 8.0 & 9.0 & gfx942 \\
        Memory (GB) & 24 & 40 & 80 & 192 \\
        BW (GB/s) & 600.1 & 1,555.2 & 3,352.3 & 4,208.7 \\
        LLC (MiB) & 6 & 40 & 50 & 256 \\
        \midrule
        FP16 peak & 16.31 & 60.55 & 107.46 & 39.09 \\
        FP32 peak & 16.31 & 15.14 & 53.73 & 128.15 \\
        FP64 peak & 0.25 & 7.57 & 26.86 & 104.97 \\
        \midrule
        FP16 BP & 27.18 & 38.93 & 32.06 & 9.29 \\
        FP32 BP & 27.18 & 9.73 & 16.03 & 30.45 \\
        FP64 BP & 0.42 & 4.87 & 8.01 & 24.94 \\
    \end{tabular}
    \par\vspace{0.35em}\raggedright\footnotesize Ceilings come from the vendor profilers: the Nsight Compute device attributes on NVIDIA, and the ROCm Compute Profiler standalone Roofline microbenchmark on MI300X. MI300X lists its ISA target in place of a compute capability.
\end{subtable}

\end{table*}

\subsection{Kernel Family Selection}
\label{sec:appendix-kernel-selection}

We selected the five kernel families in four stages, using only profiled counters and no LLM predictions.
A case here is one target-kernel cell: a family's target kernel under one variant, runtime, and GPU, excluding the \Rvariant variant because renaming leaves its counters unchanged.

\paragraph{Candidates.}
Few programs have CUDA, HIP, OpenMP-offload, and OpenCL implementations at once: \hecbench supplies the first three but no OpenCL, so OpenCL implementations came from Rodinia or were ported for this study.
Within that constraint we profiled eight candidate families on all four GPUs: \texttt{bfs}, \texttt{inversek2j}, \texttt{lavaMD}, \texttt{lud}, \texttt{particlefilter}, \texttt{particles}, \texttt{srad}, and \texttt{threadfence}.

\paragraph{Eligibility.}
Every candidate had to have all nine GPU--runtime combinations for the \Ovariant variant, a dominant-precision FLOP count that does not depend on which NVIDIA GPU ran the same source and input, any cross-runtime arithmetic-intensity difference above $2\times$ explained by a documented mechanism, and clean measurements.
Two families failed the FLOP-stability requirement.
\texttt{threadfence}'s counted FLOPs include contention-dependent atomic retries, so its dominant-precision FLOP count spreads by 37\% across the A10, A100, and H100, leaving no machine-independent value to predict; every other family spreads by at most $2\times10^{-5}$.
\texttt{bfs} is integer-only, so it has no meaningful arithmetic intensity, and its few counted FLOPs also vary across GPUs.
Cross-runtime divergence was not a reason for exclusion: \texttt{inversek2j}'s $10.9\times$ arithmetic-intensity spread across MI300X runtimes comes from OpenCL's heavier \texttt{acos} and \texttt{sqrt} library implementations, which we confirmed in the disassembly, and we treat it as a deliberately hard case.

\paragraph{Structural constraints.}
A valid set must contain a saturated compute-bound anchor and a saturated bandwidth-bound anchor, and at least one member whose regime changes along two or more of the GPU, runtime, and variant axes.
Only \texttt{lavaMD} qualifies as the compute-bound anchor and only \texttt{particlefilter} as the bandwidth-bound anchor, which leaves four valid five-family sets.

\paragraph{Ranking.}
We ranked the valid sets by the share of their cases within 1.5 octaves of the relevant balance point (\autoref{tab:kernel-selection-sets}), because accuracy far from a balance point measures an easier classification problem.
We also recorded the CB/BB balance and the number of member pairs that share both a work shape (fixed-trip, data-dependent, or single-pass) and a memory-access pattern (streaming, tiled through shared memory, or indirect).
The selected set ranks first, and it is the only valid set with no such duplicate pair (\texttt{lavaMD} and \texttt{lud} share both properties).
Re-running the ranking under eight alternative choices (mean instead of median distance, near-ridge thresholds of 1.0 and 2.0 octaves, including the \Rvariant variant, and dropping each GPU in turn) keeps the same set first in six; the 2.0-octave threshold and dropping the H100 favor replacing \texttt{srad} with \texttt{lud}.
Counter noise does not affect the ranking: no case's regime lies within three standard deviations of its balance point.
No eligible family exercises atomic operations, so the study does not cover atomic-heavy kernels.

\begin{table*}[t]
    \centering
    \caption{Structurally valid five-family sets, ordered by near-ridge share (cases within 1.5 octaves of the balance point). Median octaves is the median $|\log_2(\rai/\text{balance point})|$; balance is $\min(\cbound,\bbound)$ divided by the number of cases; collisions count member pairs with the same work shape and memory-access pattern. The selected set is in bold.}
    \label{tab:kernel-selection-sets}
    \small
    \begin{tabular}{lrrrrr}
\toprule
Five-family set & Cells & Near-ridge (\%) & Median octaves & Balance & Collisions \\
\midrule
\textbf{inversek2j, lavaMD, particlefilter, particles, srad} & 134 & 31 & 3.12 & 0.485 & 0 \\
inversek2j, lavaMD, lud, particlefilter, particles & 134 & 27 & 3.11 & 0.418 & 1 \\
inversek2j, lavaMD, lud, particlefilter, srad & 135 & 25 & 3.12 & 0.415 & 1 \\
lavaMD, lud, particlefilter, particles, srad & 134 & 25 & 3.14 & 0.388 & 1 \\
\bottomrule
\end{tabular}

\end{table*}

\subsection{Agent Tool Use and Unfinished Agent Analysis}
\label{sec:appendix-agent-behavior}

\begin{figure}[t]
    \centering
    \includegraphics[width=0.9\linewidth]{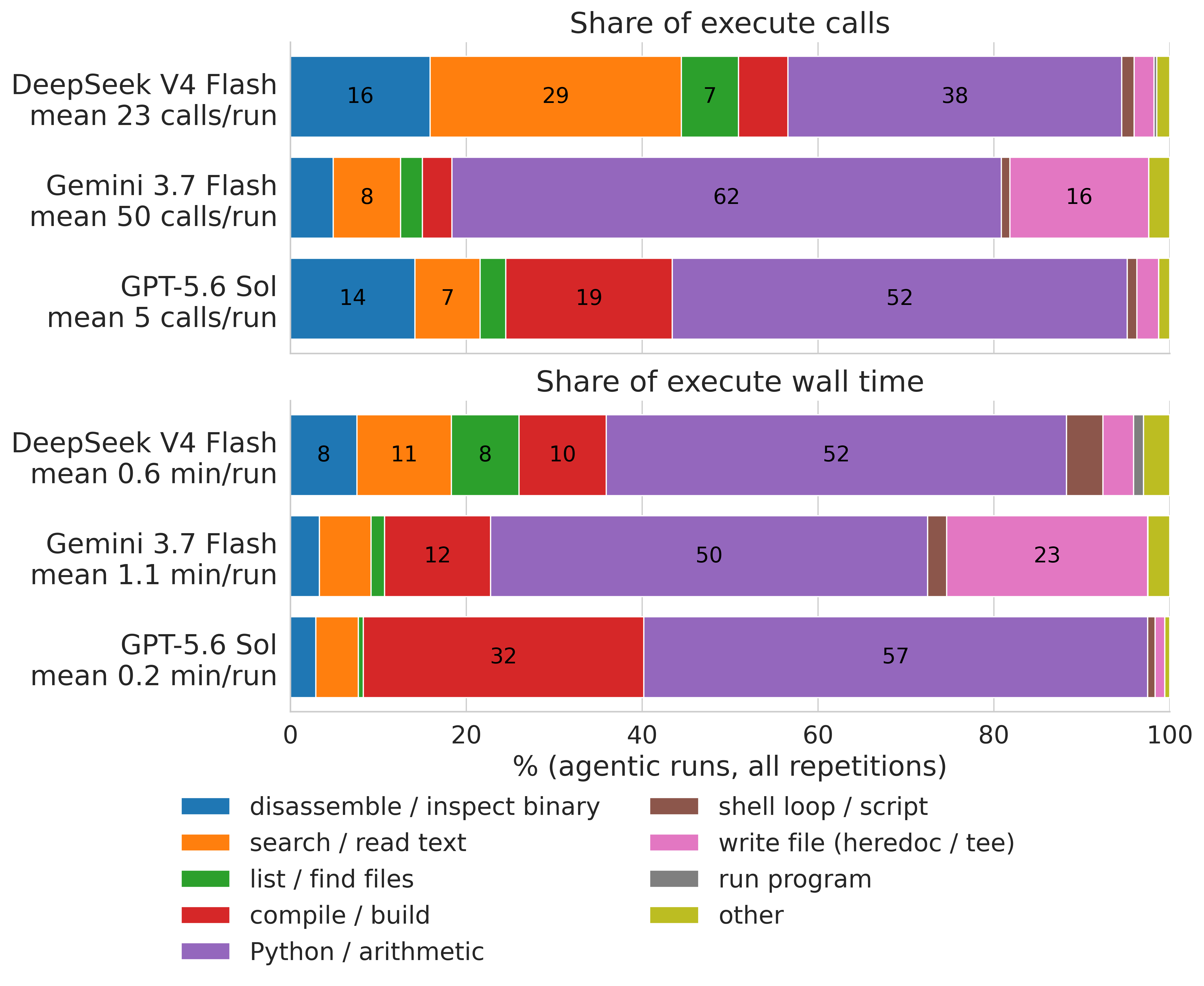}
    \caption{Commands issued through \texttt{execute}: share of calls (top) and summed call durations (bottom), pooled across all runs and repetitions.
    Each call is categorized by its first substantive command, with its full duration assigned to that category.
    Durations exclude model inference and other tools.}
    \label{fig:agent_execute_commands}
\end{figure}

\autoref{fig:agent_execute_commands} breaks down the commands issued through \texttt{execute} and their recorded durations across all agentic runs and repetitions.
Python/arithmetic commands account for 38\%, 62\%, and 52\% of execute calls for \deepseekflash, \geminiflash, and \gptsol, respectively, and approximately 50--57\% of summed execute-call durations across the models.
This category includes scripts that parse and print disassembly, tally executed instructions, and calculate candidate component counters.
Compilation and build commands account for 32\% of \gptsol's summed execute-call duration, compared with 10\% for \deepseekflash and 12\% for \geminiflash.
The mean summed execute-call duration is 0.6 minutes per run for \deepseekflash, 1.1 minutes for \geminiflash, and 0.2 minutes for \gptsol.
These durations exclude model inference and other tool calls.

\begin{figure}[t]
    \centering
    \includegraphics[width=\linewidth]{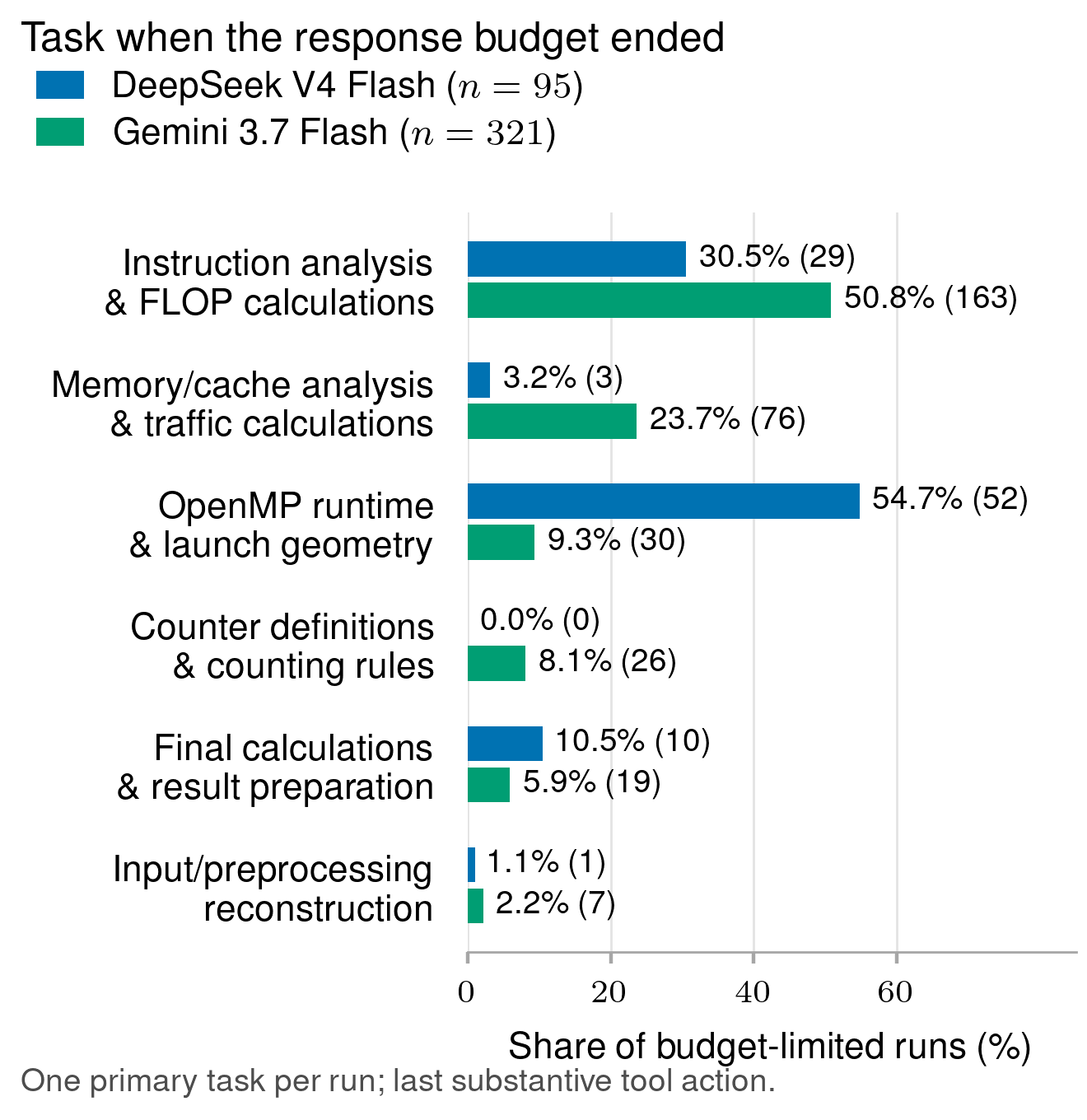}
    \caption{Primary task at budget exhaustion for 95 \deepseekflash runs and 321 \geminiflash runs, normalized separately for each model.
    Each run receives one qualitative label from its last substantive tool action, using preceding actions and generated code for context.
    Labels show percentages and run counts.}
    \label{fig:budget_tasks}
\end{figure}

\autoref{fig:budget_tasks} compares unfinished tasks in the 95 \deepseekflash runs and 321 \geminiflash runs that exhausted the 100-response budget.
Percentages are normalized separately by each model's number of budget-limited runs.
A manual review assigns each run one primary task from its last substantive tool action, using the preceding four substantive actions and earlier generated code for context.
Empty responses and task-list updates are excluded when identifying that action.
For \geminiflash, instruction analysis and FLOP calculations account for 163 runs (51\%), followed by memory/cache analysis and device-memory traffic calculations in 76 runs (24\%).
The memory/cache category includes footprint calculations, access enumeration, cache-residency analysis, and cache-model development or execution.
For \deepseekflash, OpenMP runtime and launch-geometry inspection accounts for 52 runs (55\%), followed by instruction analysis and FLOP calculations in 29 runs (31\%).
The remaining categories cover counter definitions, final calculations and result preparation, and input/preprocessing reconstruction.
These labels describe where analysis ended; they do not measure whole-run time allocation or establish which task caused non-completion.

\subsection{Completion Reliability and Resource Distributions}
\label{sec:appendix-resources}

\begin{figure}[t]
    \centering
    \includegraphics[width=\linewidth]{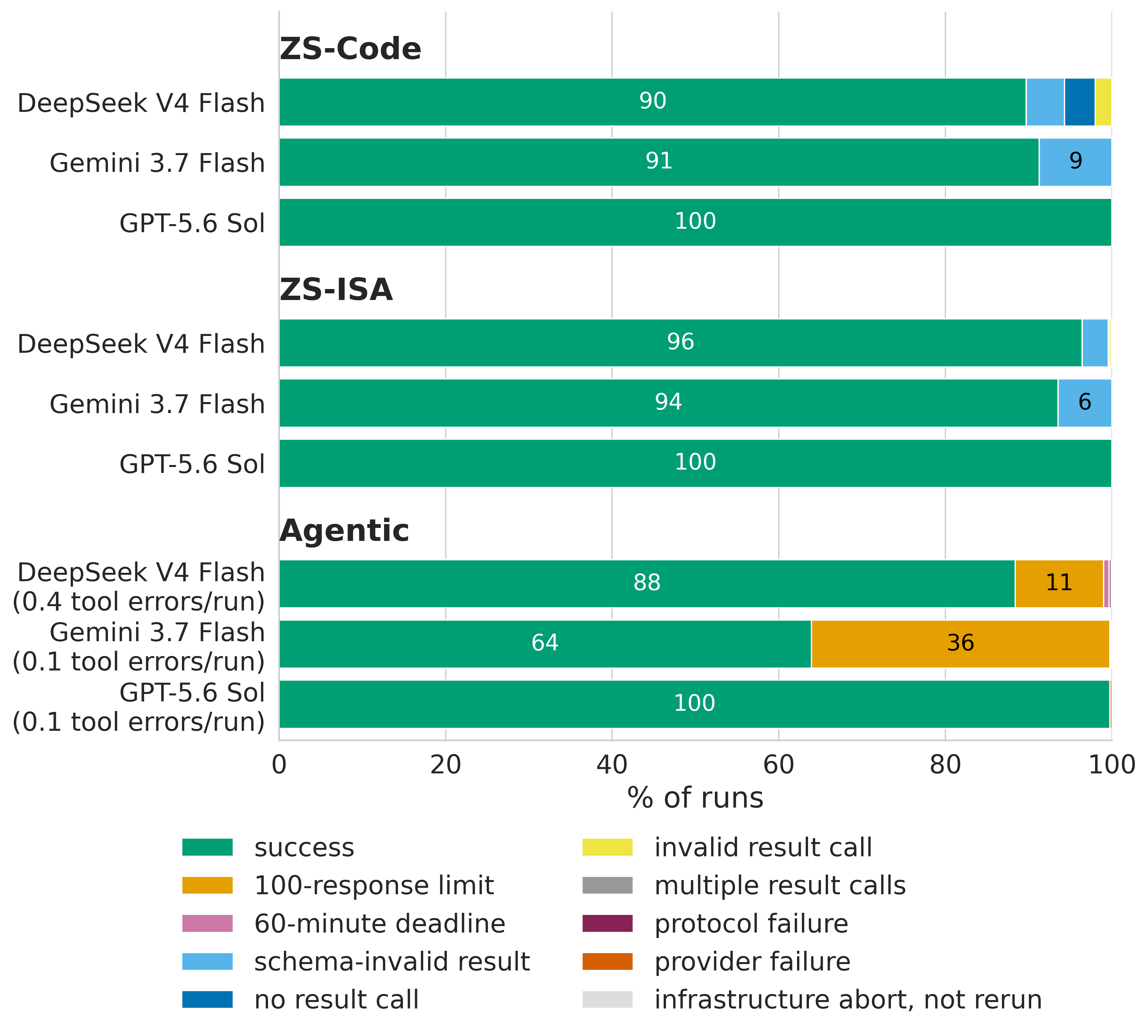}
    \caption{Run outcomes across all 895 runs per model and prompting strategy.
    Success denotes a usable structured prediction, independently of its numerical accuracy; the remaining categories distinguish termination and result-format failures.}
    \label{fig:run_outcomes}
\end{figure}

\autoref{fig:run_outcomes} separates usable predictions from unsuccessful run outcomes across all prompting strategies.
A usable prediction is a valid structured result, independently of whether its predicted counters or derived \rai are accurate.
For \deepseekflash and \geminiflash, response-budget exhaustion is the most frequent unsuccessful agentic outcome, whereas most unsuccessful zero-shot runs return schema-invalid results or omit the required result call.

\begin{figure}[t]
    \centering
    \includegraphics[width=\linewidth]{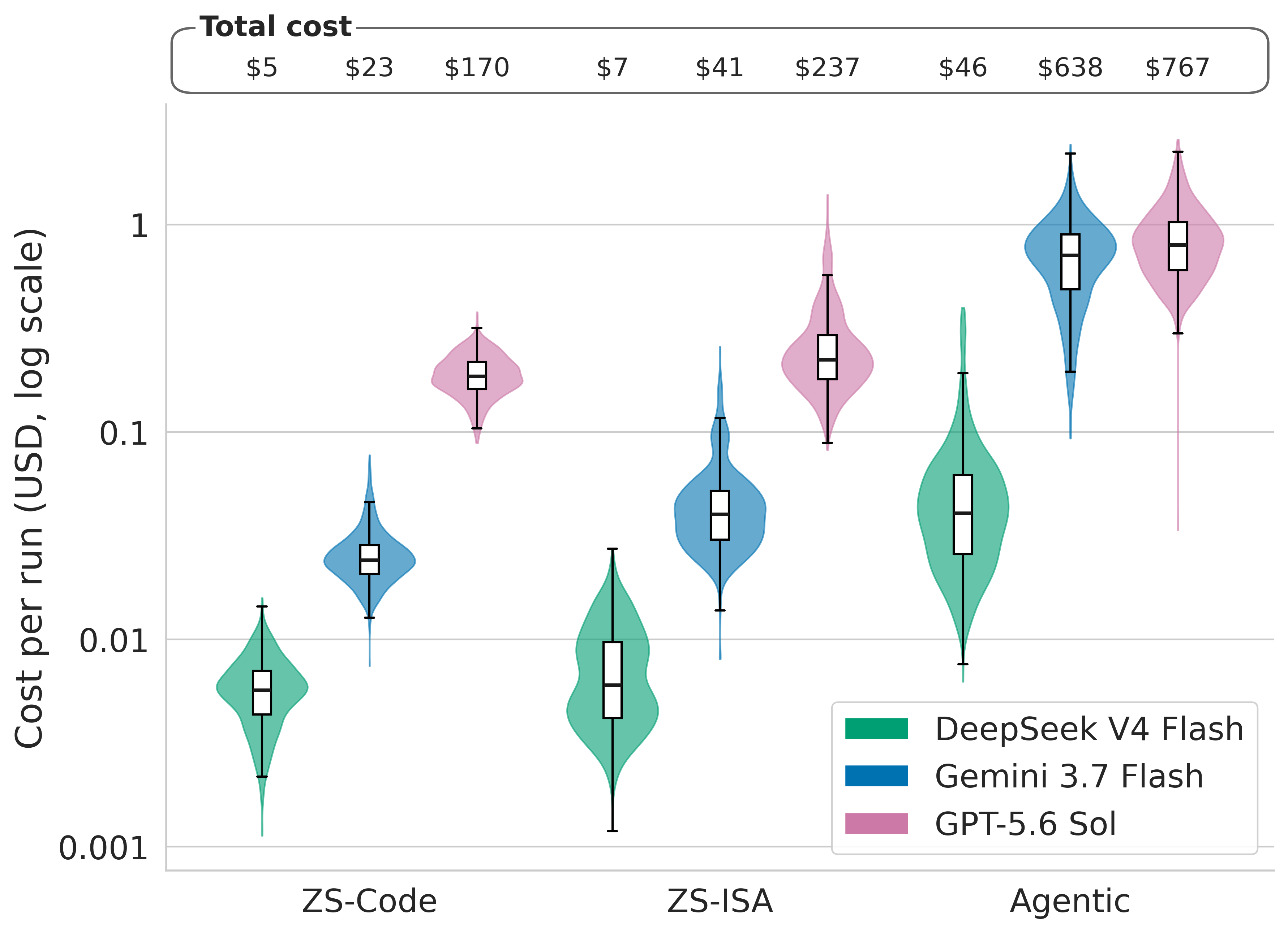}
    \caption{Recorded cost per planned run, including unsuccessful runs, on a logarithmic axis.
    Boxes show medians and interquartile ranges; the header shows summed recorded costs across 895 runs per configuration.
    Unreported charges are omitted, so affected values are lower bounds.}
    \label{fig:recorded_cost_distribution}
\end{figure}

\autoref{fig:recorded_cost_distribution} shows the distribution of recorded costs per planned run and their sums across each model--strategy combination.
Costs include recorded charges from unsuccessful submissions and earlier executions associated with the same planned run.
Unreported charges are not filled in, so affected per-run values and summed costs are lower bounds.
For \geminiflash, the 323 unsuccessful agentic runs account for \$294.51 of the \$638.24 recorded cost, or 46\%.

\begin{figure}[t]
    \centering
    \includegraphics[width=\linewidth]{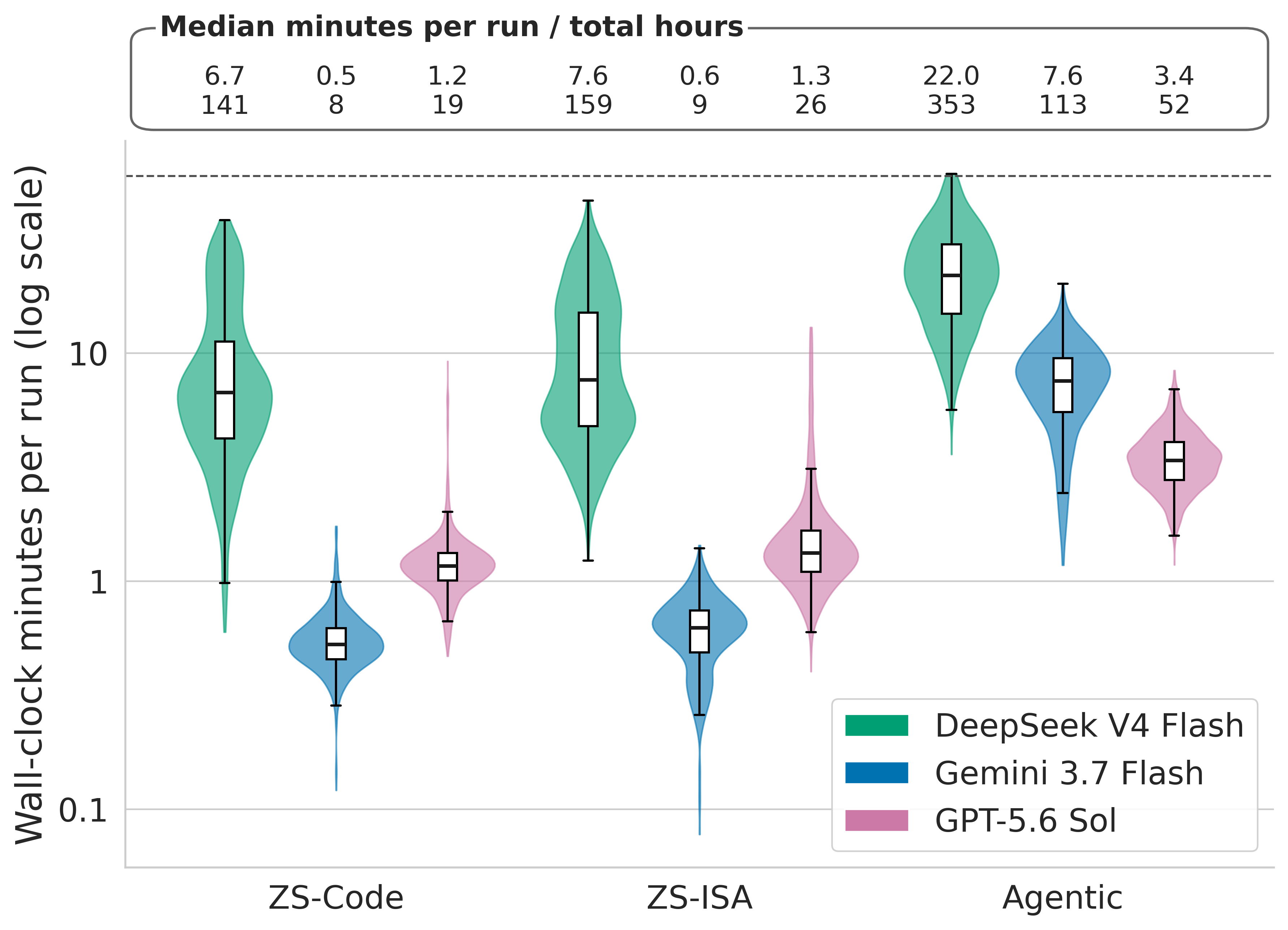}
    \caption{Per-run prediction latency across runs with recorded timings.
    Boxes show medians and interquartile ranges within the violin distributions; the header reports medians and summed run hours.
    The dashed line marks 60 minutes.
    Durations include agentic sandbox setup and analysis, excluding archival and cleanup.}
    \label{fig:prediction_latency}
\end{figure}

\autoref{fig:prediction_latency} shows the per-run prediction-latency distributions for all three prompting strategies.
For zero-shot runs, prediction latency is the accumulated recorded request-chain duration associated with a planned run.
For agentic runs, latency spans the selected execution from sandbox creation and evidence staging through model requests, tool calls, and retry waits until a prediction is submitted or execution terminates.
It excludes subsequent archival and sandbox cleanup.
One \geminiflash agentic run has no recorded duration and is omitted from the latency distribution, leaving 894 observations; the other configurations each have 895.
Summed run hours add these per-run durations and do not represent elapsed campaign time because runs can execute concurrently.

\begin{figure}[!t]
    \centering
    \includegraphics[width=\linewidth]{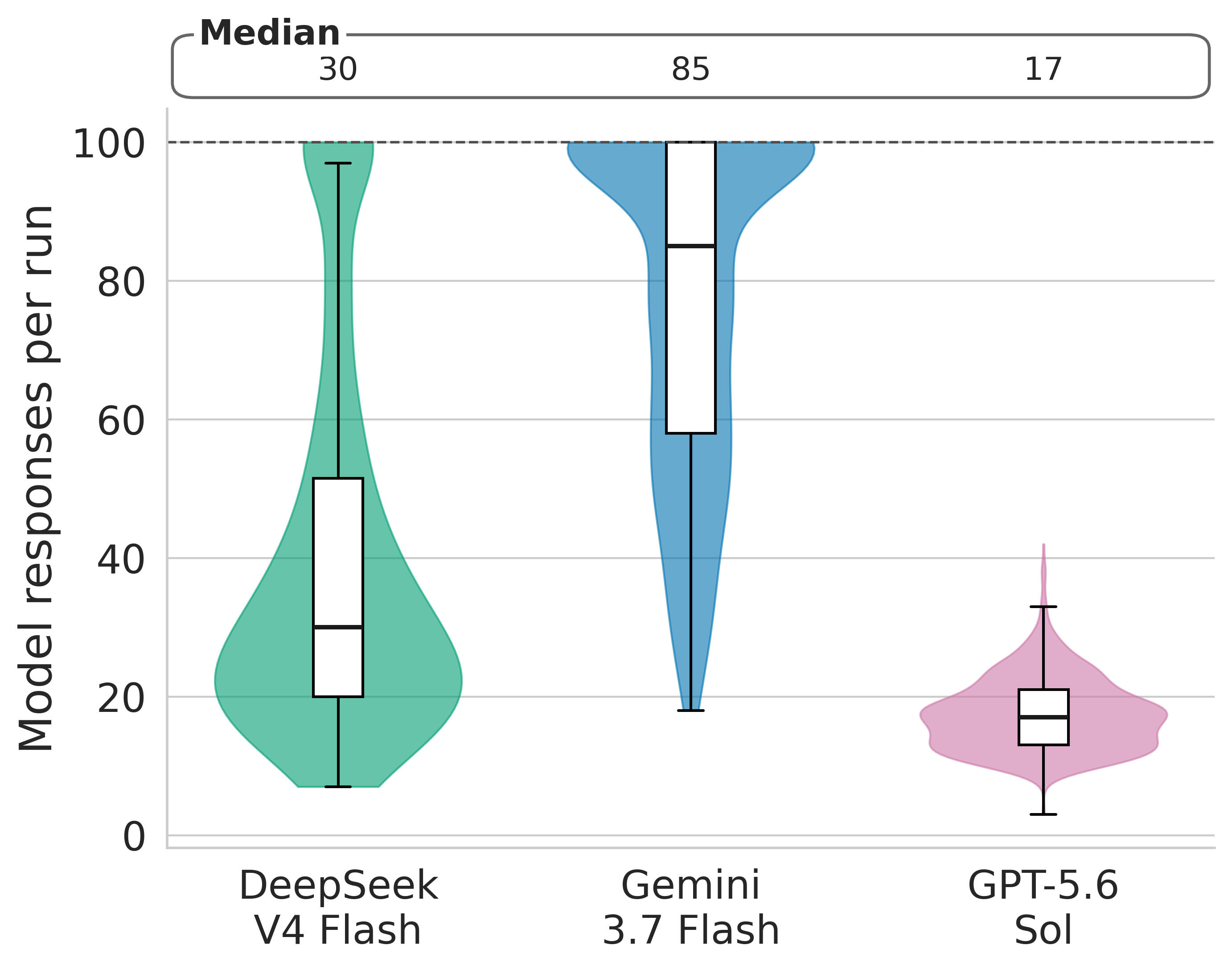}
    \caption{Delivered model responses per agentic run across all 895 runs per model.
    Boxes show medians and interquartile ranges; the dashed line marks the 100-response budget shared by analysis and summarization.}
    \label{fig:agent_model_responses}
\end{figure}

\autoref{fig:agent_model_responses} shows delivered model responses per agentic run, with medians of 30 for \deepseekflash, 85 for \geminiflash, and 17 for \gptsol.
These counts include analysis turns and any summarization responses, which share the 100-response run budget.
They differ from tool-call counts because a model response can request multiple tools or none.

\subsection{Source-Only Accuracy Across Kernels and Targets}
\label{sec:appendix-rq4}

\begin{figure*}[t]
    \centering
    \includegraphics[width=\linewidth]{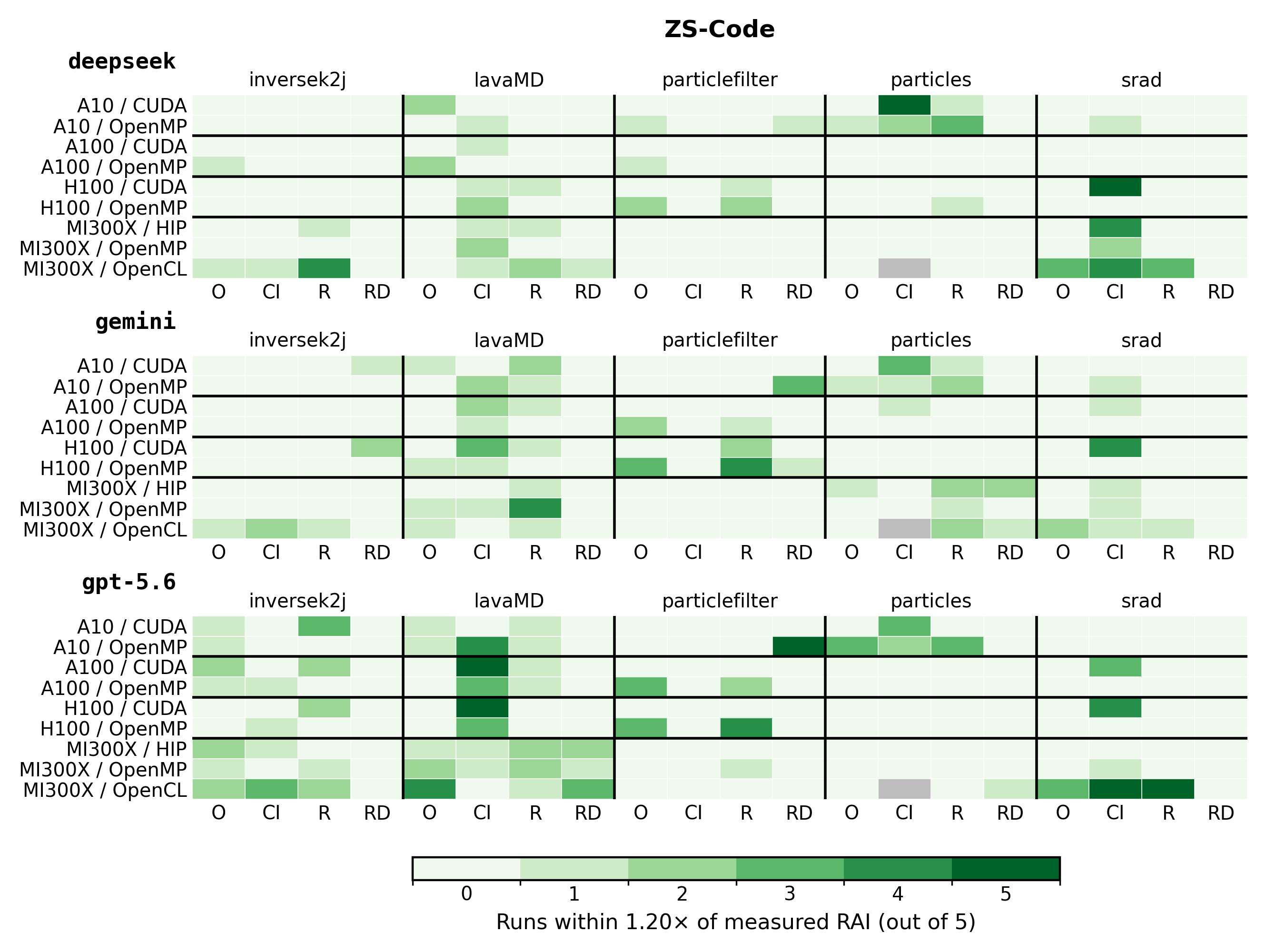}
    \caption{\rai accuracy under \zssc by kernel, variant, and GPU--runtime pair.
    Each cell counts how many of five repetitions meet the $1.20\times$ tolerance; unsuccessful predictions count as misses.
    Columns follow \Ovariant, \CIvariant, \Rvariant, and \RDvariant within each kernel.
    Gray marks the unavailable \CIvariant variant of \texttt{particles} on MI300X/OpenCL.}
    \label{fig:rq4_zs_code_accuracy}
\end{figure*}

\autoref{fig:rq4_zs_code_accuracy} provides the source-only counterpart to the joint kernel, variant, and GPU--runtime breakdown.
Each populated cell represents one case with five repetitions, including unsuccessful predictions as misses.
For \gptshort, aggregate \rai accuracy within the $1.20\times$ tolerance is 14\% on \Ovariant, 21\% on \CIvariant, 15\% on \Rvariant, and 5\% on \RDvariant.
The corresponding \zsisa accuracies are 32\%, 30\%, 34\%, and 32\%, respectively.
These comparisons show sensitivity to the supplied evidence and code variants, without establishing whether a model encountered the original kernel during training.

\subsection{Operation-Counting Conventions}
\label{sec:appendix-counter-conventions}

The prediction targets are profiler-derived counters, whose operation counts can differ from counts of mathematical operations in the source.
Our AMD reference follows VALU instruction-counter weighting: addition, multiplication, and transcendental instructions contribute one operation per work-item, while FMA contributes two, without an additional factor for packed components~\citep{amdROCmFlopConventions2026}.
NVIDIA FP16 reference counts can also include compiler-generated \texttt{HFMA2} instructions that materialize constants, even when the source has no FP16 arithmetic.
These contributions are retained in the reference targets.

\autoref{tab:inversek2j-accounting} reconciles \gptshort's \zsisa FLOP predictions for the \Ovariant and \Rvariant \texttt{inversek2j} cases on MI300X/OpenMP.
The reference total of 173 operations per work-item includes 171 in the computational body and two setup operations: a reciprocal and a multiplication.
The model instead counts both components of nine packed FMAs, five packed multiplies, and four packed adds, adding $9\times2+5+4=27$ operations relative to the reference weighting.
It also counts 12 \texttt{v\_div\_scale} and six \texttt{v\_div\_fixup} instructions as 18 further operations outside the reference arithmetic categories.
Its body estimate is therefore $171+27+18=216$, and including setup yields 218 operations per work-item, exactly matching its predicted total of 7,143,424 FLOPs over 32,768 work-items.

\begin{center}
\begin{minipage}{\linewidth}
    \captionsetup{type=table}
    \caption{Operation-count reconciliation for \gptshort's \zsisa \Ovariant and \Rvariant \texttt{inversek2j} predictions on MI300X/OpenMP, per work-item.}
    \label{tab:inversek2j-accounting}
    \centering
    \begin{tabular}{lr}
        Contribution & Operations \\
        \midrule
        Reference total, including setup & 173 \\
        Additional packed weighting & 27 \\
        Division-helper instructions & 18 \\
        \midrule
        Predicted total, including setup & 218 \\
    \end{tabular}
\end{minipage}
\end{center}

This accounting mismatch explains the $218/173\approx1.26\times$ FLOP overestimate for these cases.
Predicted total traffic is approximately $0.933\times$ the reference, so the combined \rai overestimate is about $1.35\times$.
The \RDvariant FP64 body lacks the corresponding packed FP32 arithmetic; its smaller FLOP discrepancy and more accurate traffic estimate yield \rai within the $1.20\times$ tolerance.
This reconciliation is specific to the checked model, strategy, and variants and does not attribute every MI300X error to the same convention.

\subsection{Profiling Reference Variability}
\label{sec:t-reference-variability}

Reference FLOP and byte counters are unrounded means over ten profiling trials for each of the 179 cases.
Reference \rai is the sum of the FLOP-counter means divided by the sum of the read- and write-counter means.
\autoref{tab:t-reference_variability} summarizes the per-case coefficient of variation (CoV) of the ten individual trials' \rai values, as recorded in the reference characterization.
Each trial's \rai uses its own total FLOPs and total bytes; CoV is the sample standard deviation divided by the mean, expressed as a percentage.
This describes variation across profiling trials, whereas the numerical-agreement tolerance compares an LLM estimate with the reference ratio of counter means.
Variability in one component counter need not equal variability in total traffic or \rai.

\begin{table}[tp]
    \caption{Variation in profiled \rai across ten trials per case.
    Median and Maximum summarize the per-case CoV values within each GPU group; All summarizes all 179 cases.
    These characterize profiling variability, rather than confidence intervals on LLM accuracy.}
    \label{tab:t-reference_variability}
    \centering
    \begingroup
\footnotesize
\setlength{\tabcolsep}{2pt}
\begin{tabular}{@{}lrrr@{}}
 & & \multicolumn{2}{c}{\rai CoV (\%)} \\
GPU & Cases & Median & Maximum \\
\midrule
A10 & 40 & 0.278 & 1.944 \\
A100 & 40 & 0.018 & 0.229 \\
H100 & 40 & 0.075 & 10.257 \\
MI300X & 59 & 0.056 & 0.579 \\
\midrule
All & 179 & 0.076 & 10.257 \\
\end{tabular}
\endgroup

\end{table}

\subsection{Component and Classification Accuracy}
\label{sec:t-appendix-scores}

\begin{figure}[t]
    \centering
    \includegraphics[width=0.8\linewidth]{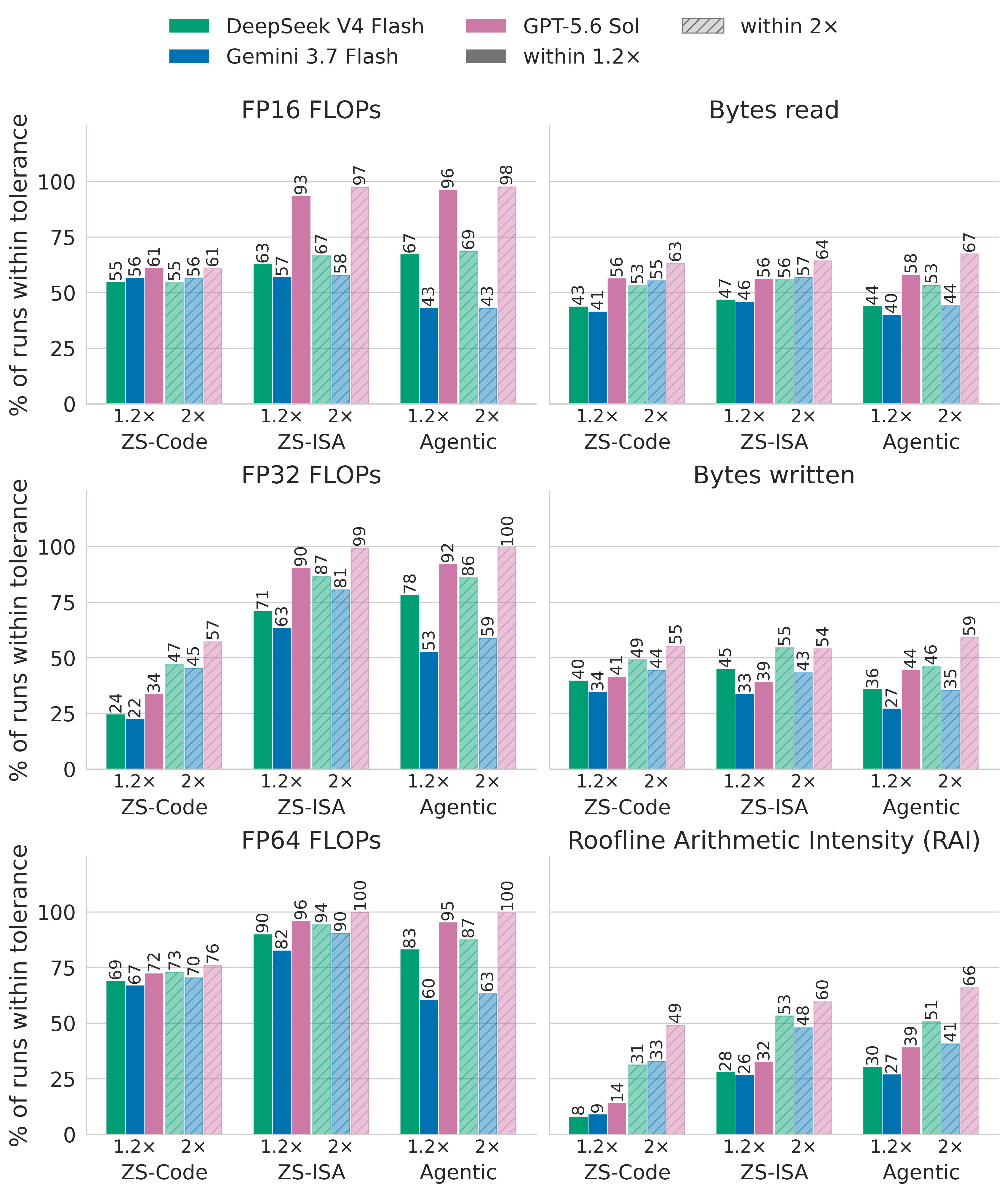}
    \caption{Accuracy of predicted component counters (FP16/32/64 FLOPs and device-memory bytes read/written) and derived \rai.
    Solid bars show the percentage of runs within the $1.20\times$ tolerance ($1/1.20$--$1.20$ of the profiled value); hatched bars use the $2\times$ tolerance ($0.5$--$2.0$).
    Failed and incomplete predictions count as misses.}
    \label{fig:t-counter_accuracy}
\end{figure}

\autoref{fig:t-counter_accuracy} separates the per-precision FLOP counters and device-memory reads and writes.
Counter accuracy uses symmetric relative error $2|\hat{x}-x|/(\hat{x}+x)$, with zero error when both quantities are zero; the $f\times$ tolerance corresponds to an error no greater than $2(f-1)/(f+1)$.
Unsuccessful predictions count as misses, including when the reference counter is zero.
\autoref{tab:t-classification_metrics} reports both reference classes and includes unclassified runs in each true class's recall denominator.
The reference set contains 93 bandwidth-bound and 86 compute-bound cases, with five repetitions per case.

\begin{table}[t]
    \caption{\bbound/\cbound classification metrics for every model and strategy (\%; 895 runs each; 93 \bbound and 86 \cbound cases).
    Unclassified runs (Uncl.: no usable prediction or non-finite \rai) count as misses for accuracy and for recall of their true class; precision is computed over runs predicting that class.
    Bal.: balanced accuracy, the mean of the two recalls.}
    \label{tab:t-classification_metrics}
    \centering
    \begingroup
\footnotesize
\setlength{\tabcolsep}{2pt}
\begin{tabular}{@{}lrrrrrrr@{}}
 & & & \multicolumn{2}{c}{\bbound} & \multicolumn{2}{c}{\cbound} & \\
Configuration & Acc. & Bal. & Prec. & Rec. & Prec. & Rec. & Uncl. \\
\midrule
\deepseekshort \zssc & 65.4 & 65.3 & 74.3 & 65.8 & 76.0 & 64.9 & 13.0 \\
\geminishort \zssc & 65.4 & 65.3 & 73.4 & 66.0 & 78.5 & 64.7 & 13.7 \\
\gptshort \zssc & 80.0 & 79.5 & 75.4 & 91.2 & 89.3 & 67.9 & 0.7 \\
\midrule
\deepseekshort \zsisa & 83.9 & 84.0 & 91.0 & 82.6 & 88.6 & 85.3 & 6.6 \\
\geminishort \zsisa & 78.3 & 78.5 & 88.0 & 74.4 & 92.0 & 82.6 & 13.0 \\
\gptshort \zsisa & 91.8 & 91.8 & 92.2 & 92.0 & 91.8 & 91.6 & 0.2 \\
\midrule
\deepseekshort \agentic & 75.3 & 75.6 & 90.6 & 68.2 & 87.7 & 83.0 & 15.4 \\
\geminishort \agentic & 53.9 & 54.3 & 91.9 & 43.9 & 91.1 & 64.7 & 41.1 \\
\gptshort \agentic & 91.8 & 92.0 & 96.0 & 88.0 & 89.0 & 96.0 & 0.6 \\
\end{tabular}
\endgroup

\end{table}

\autoref{tab:t-classification_baselines} reports a majority-class lookup using only the GPU and dominant reference precision.
We identify precision by the largest reference FLOP counter, yielding FP32 and FP64 groups on each of the four GPUs.
Within each group, the lookup predicts whichever reference class is more common; a tie defaults to \bbound without changing its in-sample accuracy.
Its 131 correct classifications out of 179 cases give 73.2\% accuracy, compared with 52.0\% for always predicting \bbound.
Both fitting and scoring use the same labels, so this is an optimistic in-sample reference for class composition rather than a held-out prediction result.

As a precision-sensitivity check, we select each prediction's balance point using its own largest FLOP counter, retaining the same hardware balance points, reference labels, and treatment of unclassified runs.
All 7,122 classifiable predictions select the same dominant precision as their references; selecting balance points from predicted precision leaves classifications unchanged across all 8,055 runs.

\begin{table}[tp]
    \caption{GPU--precision majority lookup fitted and evaluated on the same 179 cases.
    \bbound and \cbound columns give reference-case counts; Guess selects the row's majority class.
    Accuracy gives each case equal weight, equivalent to five repetitions per case; All pools the eight groups.}
    \label{tab:t-classification_baselines}
    \centering
    \begingroup
\footnotesize
\setlength{\tabcolsep}{2pt}
\begin{tabular}{@{}llrrlr@{}}
GPU & Precision & \bbound & \cbound & Guess & Acc. (\%) \\
\midrule
A10 & FP32 & 24 & 6 & \bbound & 80.0 \\
A10 & FP64 & 1 & 9 & \cbound & 90.0 \\
\midrule
A100 & FP32 & 7 & 23 & \cbound & 76.7 \\
A100 & FP64 & 2 & 8 & \cbound & 80.0 \\
\midrule
H100 & FP32 & 14 & 16 & \cbound & 53.3 \\
H100 & FP64 & 2 & 8 & \cbound & 80.0 \\
\midrule
MI300X & FP32 & 30 & 12 & \bbound & 71.4 \\
MI300X & FP64 & 13 & 4 & \bbound & 76.5 \\
\midrule
All &  & 93 & 86 & by group & 73.2 \\
\end{tabular}
\endgroup

\end{table}

\subsection{Numerical Accuracy Among Completed Predictions}
\label{sec:t-completed-accuracy}

\autoref{tab:t-completed_accuracy} separates accuracy over all runs from accuracy within each configuration's completed predictions.
Completion means a usable structured result, including predictions whose derived \rai is non-finite and therefore numerically inaccurate.
For \geminishort, 237 of 837 completed \zsisa predictions meet the $1.20\times$ tolerance, compared with 239 of 572 completed \agentic predictions: 28.3\% and 41.8\%, respectively.
The completed subsets differ between strategies and can weight cases differently, so these rates describe selected returned predictions rather than a paired improvement on the same cases.
Accuracy over all runs remains the primary end-to-end metric, with unsuccessful runs counted as misses.

\begin{table}[tp]
    \caption{Numerical \rai accuracy within $1.20\times$ over all runs and among completed predictions (895 runs per configuration).
    Complete and Accurate give counts of usable results and estimates meeting tolerance.
    Under Accuracy, All uses 895 as the denominator and Complete uses the usable-result count; completed subsets differ by strategy.}
    \label{tab:t-completed_accuracy}
    \centering
    \begingroup
\footnotesize
\setlength{\tabcolsep}{2pt}
\begin{tabular}{@{}llrrrr@{}}
 & & & & \multicolumn{2}{c}{Accuracy (\%)} \\
Model & Strategy & \shortstack{Complete\\(n)} & \shortstack{Accurate\\(n)} & All & Complete \\
\midrule
\deepseekshort & \zsisa & 863 & 248 & 27.7 & 28.7 \\
\deepseekshort & \agentic & 791 & 270 & 30.2 & 34.1 \\
\midrule
\geminishort & \zsisa & 837 & 237 & 26.5 & 28.3 \\
\geminishort & \agentic & 572 & 239 & 26.7 & 41.8 \\
\midrule
\gptshort & \zsisa & 895 & 290 & 32.4 & 32.4 \\
\gptshort & \agentic & 893 & 348 & 38.9 & 39.0 \\
\end{tabular}
\endgroup

\end{table}

\subsection{Exploratory Reliability Signals}
\label{sec:t-appendix-agreement}

We examine associations between prediction-derived signals and accuracy on the evaluation cases themselves.
These analyses are exploratory, without validation on held-out cases.
\autoref{tab:t-rq2_reliability} summarizes classification accuracy and coverage for predicted separation from the balance point and agreement across repetitions.
For a positive predicted intensity $\hat{I}$ and known hardware balance point $B$, predicted separation is $D(\hat{I})=\max(\hat{I}/B,B/\hat{I})$ and requires no target-kernel profiling.
Under \zsisa, a predicted separation greater than $4\times$ selects 46--53\% of all runs across the three models, with classification accuracy of 94--98\%.
When all five \gptshort \zsisa predictions give the same valid class, they are correct 98.6\% of the time, covering 82\% of cases.
The agreement signal requires five prediction runs per case, whereas predicted separation uses one.

\begin{table}[tbp]
    \caption{Exploratory reliability signals: single-run classification accuracy by predicted separation from the balance point (top) and five-run agreement (bottom).}
    \label{tab:t-rq2_reliability}
    \centering
    \begingroup
\footnotesize
\setlength{\tabcolsep}{3pt}
\begin{tabular}{@{}lrrrr@{}}
\multicolumn{5}{@{}l}{\textit{Single run: predicted distance from balance point}} \\
Configuration & $\le 2\times$ & $(2, 4]\times$ & $>4\times$ & All \\
\midrule
\deepseekshort \zsisa & 80 (26) & 93 (16) & 94 (52) & 84 \\
\geminishort \zsisa & 77 (26) & 89 (15) & 98 (46) & 78 \\
\gptshort \zsisa & 79 (28) & 95 (19) & 98 (53) & 92 \\
\gptshort \agentic & 81 (28) & 96 (19) & 97 (52) & 92 \\
\midrule
\multicolumn{5}{@{}l}{\textit{Five runs: agreement across repetitions}} \\
 & \multicolumn{2}{c}{Same class} & \multicolumn{2}{c}{\rai within $1.20\times$} \\
Configuration & Cases & Correct & Cases & Accurate \\
\midrule
\deepseekshort \zsisa & 110/179 & 97.3 & 12/179 & 60 / 25 \\
\geminishort \zsisa & 105/179 & 93.3 & 27/179 & 52 / 22 \\
\gptshort \zsisa & 147/179 & 98.6 & 51/179 & 55 / 23 \\
\gptshort \agentic & 149/179 & 98.0 & 59/179 & 56 / 30 \\
\end{tabular}
\endgroup

    \par\smallskip
    {\footnotesize\raggedright
    \textit{Top:} classification accuracy (\%), with the band's share of all 895 runs in parentheses.
    Bands use the predicted \rai; failed or non-finite predictions fall in no band, so shares need not sum to 100.
    \textit{All:} accuracy over all runs, with unsuccessful runs counted as misses.
    \textit{Bottom:} cases (of 179) whose five predictions agree: all five give the same valid class (\textit{Same class}), or all five \rai estimates are finite and positive with max/min $\le 1.20$ (\textit{\rai within $1.20\times$}).
    \textit{Correct:} classification accuracy within the agreeing cases.
    \textit{Accurate:} \% of predictions within $1.20\times$ of the reference in agreeing / remaining cases.\par}
\end{table}

Numerical consistency provides weaker assurance.
For \gptshort under \zsisa, 51 of 179 cases (28\%) have five finite, positive \rai estimates agreeing within $1.20\times$ of one another.
Among predictions in those cases, 55\% meet the numerical tolerance against the reference, compared with 23\% in the remaining cases.
Agreement is associated with a higher share of accurate estimates but can also reproduce a shared error.
\autoref{tab:t-agreement} extends the five-run agreement results to all nine model--strategy configurations.
Numerical accuracy is the share of individual predictions meeting the reference tolerance within the selected cases, rather than the accuracy of an aggregate prediction.
For \gptshort under \zsisa, the 51 numerically agreeing cases divide into 28 with all five predictions accurate and 23 with all five inaccurate.
Other configurations also have agreeing cases with a mixture of accurate and inaccurate predictions, as defined in the caption.
The numerical consistency association is weaker or absent under source-only prompting, so the compiler-informed results should not be interpreted as a universal calibration rule.

\begin{table}[t]
    \caption{Five-run agreement for every model and strategy; case counts are out of 179.
    Same class: all five predictions give the same valid \bbound/\cbound class; Correct: classification accuracy of those predictions.
    \rai within $1.20\times$: all five \rai estimates are finite and positive with max/min $\le 1.20$; Acc.: \% of predictions within $1.20\times$ of the reference in agreeing / remaining cases.
    In / Out: agreeing \rai cases whose five predictions are all within, or all outside, $1.20\times$ of the reference; the remaining agreeing cases contain a mixture of predictions within and outside tolerance.}
    \label{tab:t-agreement}
    \centering
    \begingroup
\footnotesize
\setlength{\tabcolsep}{2pt}
\begin{tabular}{@{}lrrrrrr@{}}
 & \multicolumn{2}{c}{Same class} & \multicolumn{4}{c}{\rai within $1.20\times$} \\
Configuration & Cases & Correct & Cases & Acc. & In & Out \\
\midrule
\deepseekshort \zssc & 104 & 77.9 & 20 & 9 / 8 & 1 & 18 \\
\geminishort \zssc & 95 & 77.9 & 10 & 2 / 9 & 0 & 9 \\
\gptshort \zssc & 144 & 84.0 & 43 & 15 / 13 & 5 & 34 \\
\midrule
\deepseekshort \zsisa & 110 & 97.3 & 12 & 60 / 25 & 7 & 4 \\
\geminishort \zsisa & 105 & 93.3 & 27 & 52 / 22 & 12 & 11 \\
\gptshort \zsisa & 147 & 98.6 & 51 & 55 / 23 & 28 & 23 \\
\midrule
\deepseekshort \agentic & 92 & 98.9 & 13 & 91 / 25 & 11 & 1 \\
\geminishort \agentic & 26 & 100.0 & 21 & 71 / 21 & 15 & 6 \\
\gptshort \agentic & 149 & 98.0 & 59 & 56 / 30 & 30 & 24 \\
\end{tabular}
\endgroup

\end{table}

\subsection{Workload, Target, and Variant Breakdowns}
\label{sec:t-appendix-breakdowns}

\begin{figure*}[t]
    \centering
    \includegraphics[width=\linewidth]{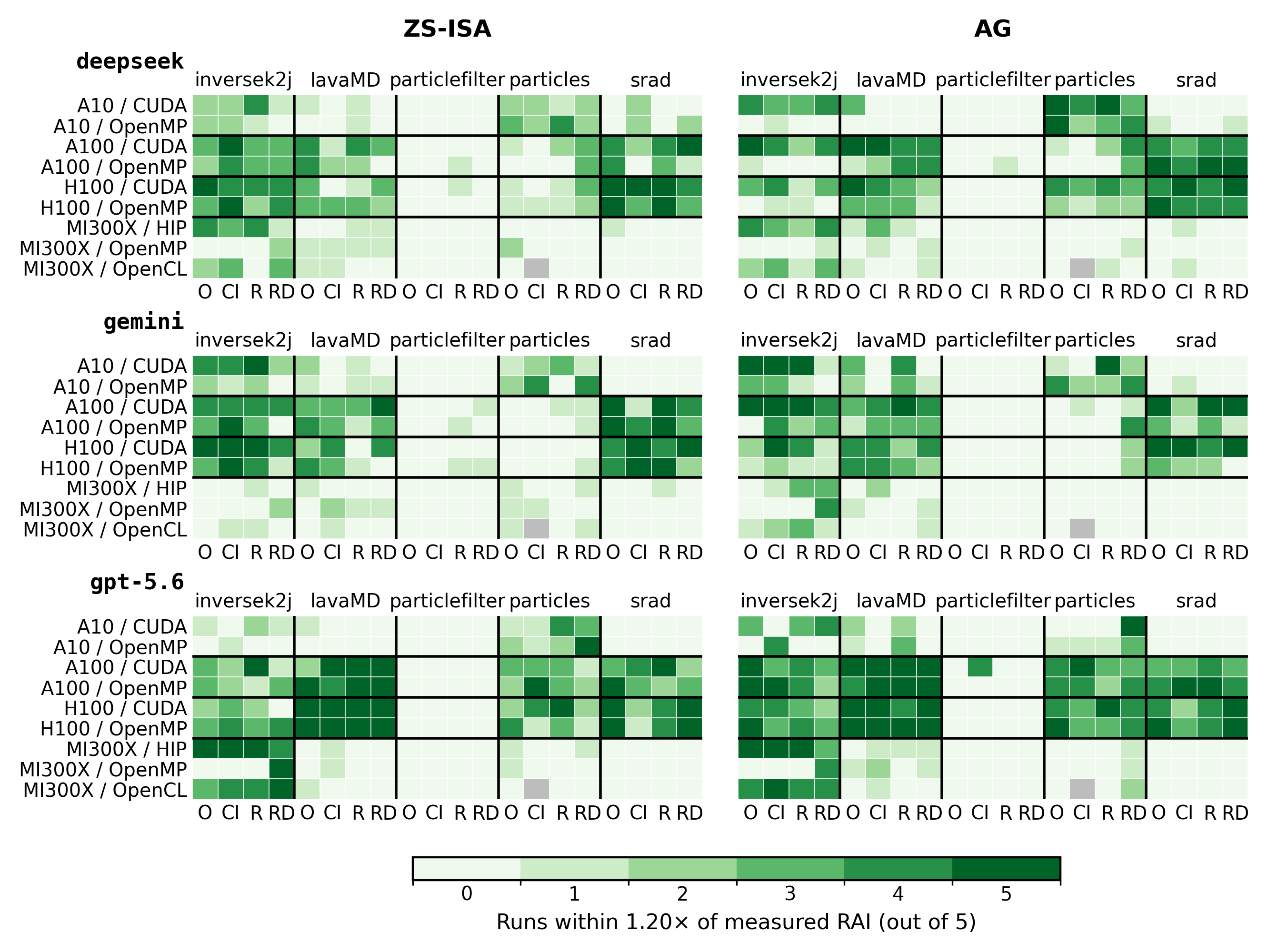}
    \caption{\rai accuracy by kernel, variant, and GPU--runtime pair under \zsisa and \agentic.
    Each cell counts how many of five repetitions meet the $1.20\times$ tolerance; unsuccessful predictions count as misses.
    Columns follow \Ovariant, \CIvariant, \Rvariant, and \RDvariant within each kernel.
    Gray marks the unavailable \CIvariant variant of \texttt{particles} on MI300X/OpenCL.}
    \label{fig:t-joint_accuracy}
\end{figure*}

\autoref{fig:t-joint_accuracy} retains each kernel, variant, and GPU--runtime combination under \zsisa and \agentic; the source-only counterpart is \autoref{fig:rq4_zs_code_accuracy}.
\autoref{tab:t-variants} summarizes numerical accuracy and usable-prediction rates across all four variants.
For \gptshort, original-versus-datatype-variant accuracy changes from 13.8\% versus 5.3\% under \zssc to 32.4\% for both under \zsisa.
The corresponding gap narrows for \deepseekshort and increases slightly for \geminishort.
Renaming produces no consistent directional accuracy change across models and strategies.
These comparisons measure sensitivity to transformations of the supplied program and do not identify whether a model encountered a kernel during training.

\begin{table}[t]
    \caption{\rai accuracy within $1.20\times$ and completion (usable prediction) by variant, for every model and strategy.
    Failed runs count as misses for accuracy; \CIvariant has fewer cases because \texttt{particles} \CIvariant is unavailable on MI300X/OpenCL.}
    \label{tab:t-variants}
    \centering
    \begingroup
\footnotesize
\setlength{\tabcolsep}{2pt}
\begin{tabular}{@{}lrrrr@{\hspace{5pt}}rrrr@{}}
 & \multicolumn{4}{c}{Within $1.20\times$ (\%)} & \multicolumn{4}{c}{Completion (\%)} \\
Configuration & \Ovariant & \CIvariant & \Rvariant & \RDvariant & \Ovariant & \CIvariant & \Rvariant & \RDvariant \\
\midrule
\deepseekshort \zssc & 6.2 & 15.0 & 8.9 & 0.9 & 97 & 94 & 95 & 73 \\
\geminishort \zssc & 6.2 & 11.8 & 12.4 & 4.4 & 90 & 94 & 92 & 89 \\
\gptshort \zssc & 13.8 & 20.9 & 15.1 & 5.3 & 100 & 100 & 100 & 100 \\
\midrule
\deepseekshort \zsisa & 30.7 & 25.0 & 28.0 & 27.1 & 99 & 100 & 99 & 88 \\
\geminishort \zsisa & 27.6 & 28.6 & 26.2 & 23.6 & 96 & 94 & 93 & 92 \\
\gptshort \zsisa & 32.4 & 30.5 & 34.2 & 32.4 & 100 & 100 & 100 & 100 \\
\midrule
\deepseekshort \agentic & 32.9 & 29.5 & 26.7 & 31.6 & 92 & 90 & 87 & 85 \\
\geminishort \agentic & 24.9 & 26.4 & 28.9 & 26.7 & 54 & 75 & 65 & 62 \\
\gptshort \agentic & 39.1 & 39.1 & 36.9 & 40.4 & 100 & 99 & 100 & 100 \\
\end{tabular}
\endgroup

\end{table}

\subsection{Representative Device-Memory Traffic Cases}
\label{sec:t-appendix-cases}

\autoref{tab:t-cases}(a) compares the original-variant CUDA implementations on NVIDIA with HIP on MI300X.
Each row reports the reference total traffic, the median prediction across five repetitions, and the number of individual \rai estimates within tolerance.
A median traffic ratio close to one does not imply that every individual \rai estimate is accurate, since each run has its own FLOP and traffic estimates.

For \texttt{lavaMD}, repeated accesses to neighboring boxes offer reuse, while the amount of traffic crossing the measured boundary varies substantially across GPUs.
For the original CUDA cases, profiled total traffic is 149--152\,MB on A100/H100 and 290\,MB on A10.
The median predicted-to-profiled ratios are 0.92--0.93 and 0.67, respectively.
This contrast is consistent with greater effective reuse on A100/H100, without isolating cache capacity, scheduling, or replacement as a sole cause.

For the original CUDA \texttt{particlefilter}, 400,000 particles each generate 69 integer indices, yielding a 110.4\,MB temporary array.
At a fixed loop iteration, adjacent threads access entries 276 bytes apart.
The kernel writes and subsequently reads this array, making traffic sensitive to transaction granularity and effective reuse.
Median predicted total traffic is approximately 4--26$\times$ below the profiled totals across the four CUDA/HIP cases, and none of these \rai estimates meets the tolerance.
The \texttt{particles} case further illustrates the limits of footprint-based estimates: MI300X records 4.8\,MB of traffic against a median prediction of 0.8\,MB despite the small data footprint.
Nominal capacity alone does not establish which requests cross the counter's memory boundary.

\begin{table}[t]
    \caption{Representative \gptshort traffic cases.
    (a) \gptshort \zsisa on the \Ovariant variant (CUDA on NVIDIA, HIP on MI300X): profiled total device-memory traffic, median predicted traffic over five runs, their ratio, and runs with \rai within $1.20\times$.
    MI300X has eight XCD-local 4\,MiB L2 caches and a 256\,MB Infinity Cache.
    (b) \gptshort \agentic \texttt{lavaMD} runs (all variants and runtimes) whose tool-call arguments contain LRU cache-simulator code patterns (Detected); a pattern match does not confirm that the simulator ran, and other implementations may go undetected.
    \rai accuracy within $1.20\times$ for runs with patterns detected / not detected is observational, since agents choose when to simulate.}
    \label{tab:t-cases}
    \centering
    \begingroup
\footnotesize
\setlength{\tabcolsep}{3pt}
\begin{tabular}{@{}lrrrrr@{}}
\multicolumn{6}{@{}l}{\textit{(a) \zsisa, \Ovariant variant: total traffic}} \\
GPU & L2 (MiB) & Prof.\ (MB) & Pred.\ (MB) & Ratio & \rai in \\
\midrule
\multicolumn{6}{@{}l}{\texttt{lavaMD}} \\
A10 & 6 & 290 & 195 & 0.67 & 1/5 \\
A100 & 40 & 152 & 141 & 0.93 & 2/5 \\
H100 & 50 & 149 & 137 & 0.92 & 5/5 \\
MI300X & $8\times4$ & 814 & 547 & 0.67 & 0/5 \\
\midrule
\multicolumn{6}{@{}l}{\texttt{particlefilter}} \\
A10 & 6 & 2,198 & 117 & 0.05 & 0/5 \\
A100 & 40 & 323 & 82 & 0.26 & 0/5 \\
H100 & 50 & 370 & 71 & 0.19 & 0/5 \\
MI300X & $8\times4$ & 3,231 & 123 & 0.04 & 0/5 \\
\midrule
\multicolumn{6}{@{}l}{\texttt{particles}} \\
A10 & 6 & 0.9 & 0.8 & 0.84 & 1/5 \\
A100 & 40 & 0.8 & 0.8 & 0.98 & 3/5 \\
H100 & 50 & 0.8 & 1.0 & 1.32 & 2/5 \\
MI300X & $8\times4$ & 4.8 & 0.8 & 0.16 & 1/5 \\
\midrule
\multicolumn{6}{@{}l}{\textit{(b) \agentic \texttt{lavaMD}: LRU simulator code}} \\
GPU & & Detected & \multicolumn{3}{r}{\rai acc.\ (\%) detected / not} \\
\midrule
A10 & & 15/40 & \multicolumn{3}{r}{33 / 12} \\
A100 & & 10/40 & \multicolumn{3}{r}{100 / 97} \\
H100 & & 5/40 & \multicolumn{3}{r}{100 / 97} \\
MI300X & & 26/60 & \multicolumn{3}{r}{12 / 15} \\
\end{tabular}
\endgroup

\end{table}

\subsection{Cache-Simulator Code in Agentic Trajectories}
\label{sec:t-cache-sim}

\begin{figure}[t]
    \centering
    \includegraphics[width=0.9\linewidth]{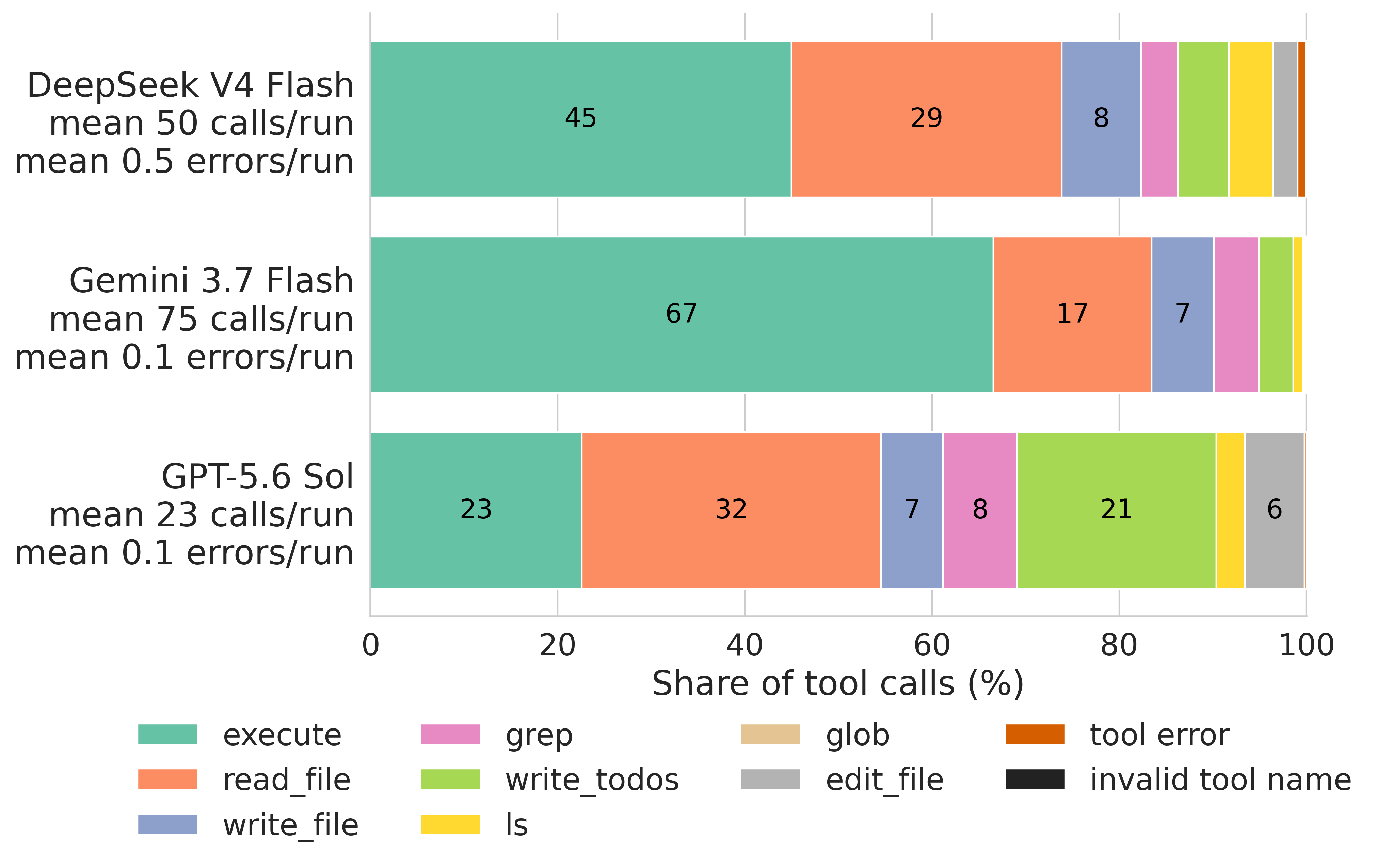}
    \caption{Distribution of recorded agent tool calls across all runs and repetitions, including unsuccessful runs.
    Labels show mean calls and errors per run; tool errors and invalid tool names form separate categories.}
    \label{fig:t-tool_calls}
\end{figure}

Inspected \gptshort \agentic \texttt{lavaMD} trajectories include Python LRU cache simulators over neighboring boxes, including an implementation modeling MI300X's separate XCD-local L2 caches.
For a reproducible code-pattern summary, we flag tool-call arguments containing \texttt{move\_to\_end}, \texttt{OrderedDict()}, or \texttt{popitem(last=False)}.
These patterns occur in 56 of 180 \texttt{lavaMD} runs, including 15 of 40 on A10 and 26 of 60 on MI300X (\autoref{tab:t-cases}(b)).
Pattern matches identify candidate simulator code; they do not confirm successful execution, and other implementations may go undetected.
The table compares accuracy in runs with and without these patterns, but agents choose when to simulate, so the comparison establishes neither a causal benefit nor its absence.
\autoref{fig:t-tool_calls} provides the overall tool-call distribution.

\subsection{Uncertainty in Accuracy Differences}
\label{sec:t-bootstrap}

\autoref{tab:t-accuracy_differences} reports paired case-bootstrap intervals for the numerical improvements from compiler evidence (RQ1), \gptshort's near-boundary classification improvement (RQ2), and the numerical differences from sandboxed analysis (RQ4).
RQ1 and RQ4 use all 179 cases; RQ2 uses the 40 cases whose profiled \rai lies within $2\times$ of the balance point.
Each of 20,000 resamples draws cases with replacement, keeping each case's five repetitions under both compared strategies together, with unsuccessful runs retained as misses.
We calculate the mean paired accuracy difference for each resample and report the 2.5th and 97.5th percentiles in percentage points, using random seed zero in \texttt{paper\_figures.py}.
The paired unit is the case, so repetitions are not treated as independent cases.
Cases also share variants, runtimes, and kernel families; these intervals describe variation over the evaluated case set and do not establish generalization beyond the five families.

\begin{table}[tp]
    \caption{Accuracy differences and paired case-bootstrap 95\% intervals for the evaluation's headline comparisons.
    Gain and interval are in percentage points, with positive values favoring the first strategy in Difference.
    Numerical accuracy uses the $1.20\times$ tolerance; intervals use 20,000 resamples retaining all five repetitions per case and strategy, including unsuccessful runs.}
    \label{tab:t-accuracy_differences}
    \centering
    \begingroup
\footnotesize
\setlength{\tabcolsep}{2pt}
\begin{tabular}{@{}llrrr@{}}
Model & Difference & Cases & Gain & 95\% interval \\
\midrule
\multicolumn{5}{@{}l}{\textit{RQ1: numerical \rai within $1.20\times$}} \\
\deepseekshort & \zsisa $-$ \zssc & 179 & 20.0 & $[14.5,25.5]$ \\
\geminishort & \zsisa $-$ \zssc & 179 & 17.8 & $[12.2,23.6]$ \\
\gptshort & \zsisa $-$ \zssc & 179 & 18.7 & $[12.0,25.4]$ \\
\midrule
\multicolumn{5}{@{}l}{\textit{RQ2: classification, profiled $D(I)\leq2$}} \\
\gptshort & \zsisa $-$ \zssc & 40 & 37.0 & $[19.5,54.0]$ \\
\midrule
\multicolumn{5}{@{}l}{\textit{RQ4: numerical \rai within $1.20\times$}} \\
\deepseekshort & \agentic $-$ \zsisa & 179 & 2.5 & $[-1.8,6.6]$ \\
\geminishort & \agentic $-$ \zsisa & 179 & 0.2 & $[-3.4,3.7]$ \\
\gptshort & \agentic $-$ \zsisa & 179 & 6.5 & $[3.4,9.6]$ \\
\end{tabular}
\endgroup

\end{table}

\subsection{Sensitivity to Numerical Tolerance}
\label{sec:t-tolerance}

\textbf{Compiler-evidence gains at tighter tolerances and \gptshort's agentic advantage extend beyond the primary $1.20\times$ target.}
Figure~\ref{fig:t-tolerance} shows the share of runs meeting numerical tolerances from $1\times$ to $5\times$ for \rai, total FLOPs, and total device-memory traffic.
\autoref{tab:t-tolerance} gives \rai accuracy at six selected factors for every model and strategy.

At $1.10\times$, $1.20\times$, $1.50\times$, and $2\times$, \zsisa improves numerical \rai accuracy over \zssc for all three models.
For \gptshort, observed \agentic accuracy remains above \zsisa throughout the $1.10\times$--$5\times$ range.
Coarser tolerances narrow the source-only versus compiler-informed difference for \gptshort: at $4\times$, accuracy is 77.0\% under \zssc and 76.5\% under \zsisa, and both reach 80.6\% at $5\times$.

For \deepseekshort and \geminishort, the small agentic differences at $1.20\times$ become lower accuracy than \zsisa at $1.50\times$, $2\times$, $4\times$, and $5\times$, reinforcing the model dependence of sandboxing's benefits.
Here, the tolerance factor compares a prediction with its profiled numerical reference; the distance bands in RQ2 instead compare profiled \rai with the hardware balance point.

\begin{table*}[p]
    \centering
    \begin{minipage}{\linewidth}
    \centering
    \includegraphics[width=0.9\linewidth]{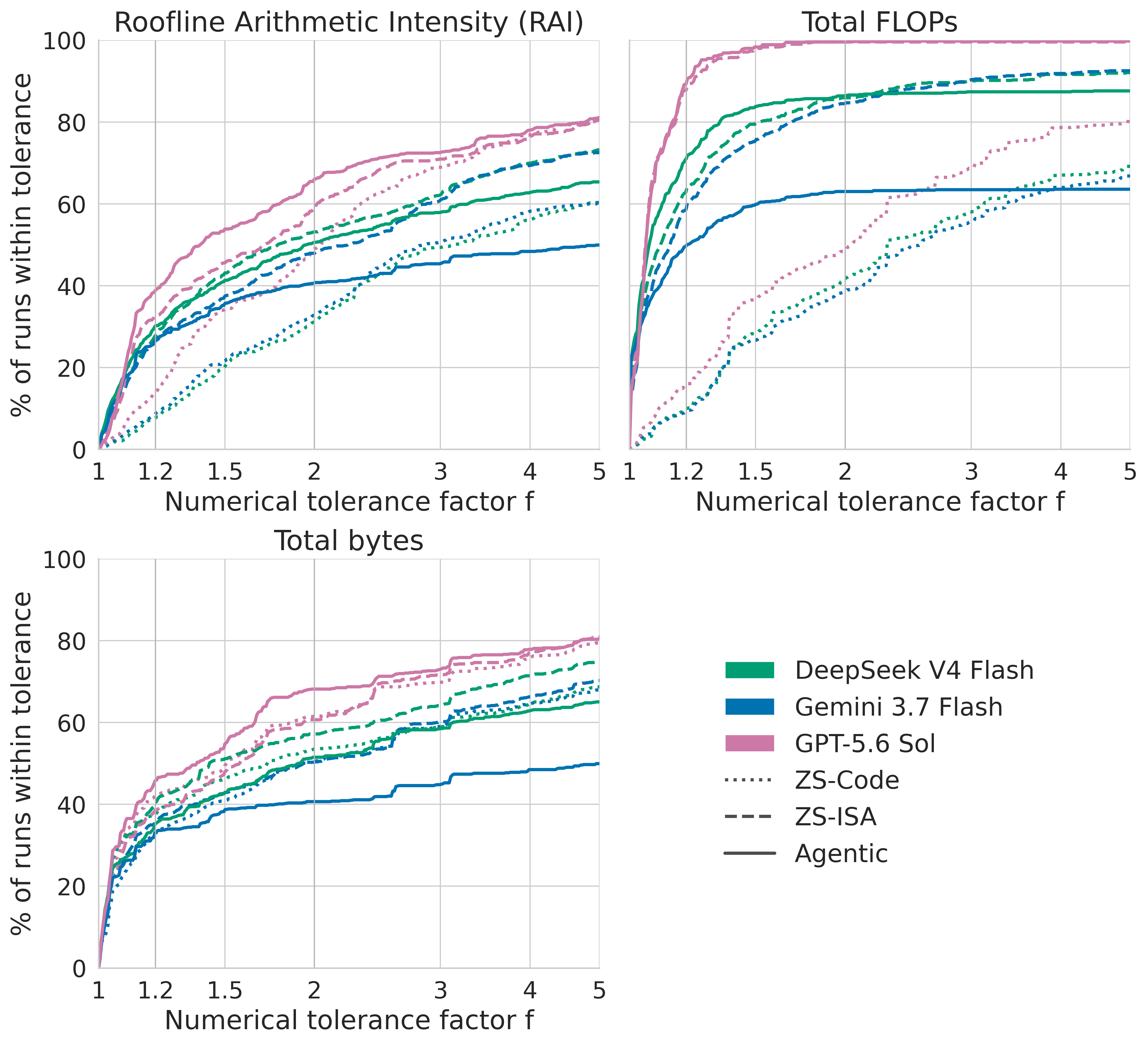}
    \captionof{figure}{Compiler evidence improves numerical accuracy at tighter tolerances, while coarser tolerances narrow some differences.
    Panels show \rai, total FLOPs, and total device-memory traffic as the numerical tolerance factor varies from 1 to 5 on a logarithmic axis; higher curves indicate more accurate predictions at that tolerance.
    Colors identify models and line styles identify strategies; vertical guides mark $1.20\times$ and $2\times$.
    Every curve includes all 895 runs per configuration, with failed and incomplete predictions counted as misses.}
    \label{fig:t-tolerance}
    \end{minipage}
    \par\medskip
    \caption{\rai accuracy (\%) at selected numerical tolerance factors for all nine model--strategy configurations.
    Each entry includes all 895 runs, with failed and incomplete predictions counted as misses; percentages are rounded to one decimal place.}
    \label{tab:t-tolerance}
    \centering
    \begingroup
\footnotesize
\setlength{\tabcolsep}{2pt}
\begin{tabular}{@{}lrrrrrr@{}}
Configuration & $1.10\times$ & $1.20\times$ & $1.50\times$ & $2\times$ & $4\times$ & $5\times$ \\
\midrule
\deepseekshort \zssc & 3.4 & 7.7 & 20.2 & 31.2 & 56.1 & 60.3 \\
\geminishort \zssc & 4.5 & 8.7 & 21.7 & 32.8 & 58.2 & 60.3 \\
\gptshort \zssc & 7.5 & 13.7 & 34.2 & 49.1 & 77.0 & 80.6 \\
\midrule
\deepseekshort \zsisa & 18.2 & 27.7 & 43.1 & 53.2 & 69.9 & 73.2 \\
\geminishort \zsisa & 17.0 & 26.5 & 37.4 & 47.9 & 69.4 & 72.5 \\
\gptshort \zsisa & 20.7 & 32.4 & 45.6 & 59.7 & 76.5 & 80.6 \\
\midrule
\deepseekshort \agentic & 20.3 & 30.2 & 41.3 & 50.6 & 62.8 & 65.4 \\
\geminishort \agentic & 18.2 & 26.7 & 35.6 & 40.7 & 48.4 & 49.9 \\
\gptshort \agentic & 23.9 & 38.9 & 53.9 & 66.0 & 78.0 & 81.1 \\
\end{tabular}
\endgroup

\end{table*}

\end{document}